\documentclass[preprint,12pt]{elsarticle}
\usepackage{amssymb}
\usepackage{mathtools}
\usepackage{amsmath}
\usepackage{amsthm}
\usepackage[thinc]{esdiff}
\usepackage{subcaption}
\usepackage{color, soul}
\usepackage{xcolor}
\usepackage{cleveref}

\theoremstyle{definition}

\usepackage{xcolor}
\usepackage{tikz}
\usetikzlibrary{calc}
\usepackage{tikz,pgfplots}
\usetikzlibrary{positioning}
\usepackage{amsmath,amsthm,amssymb, tikz, graphicx}
\usepackage[thinc]{esdiff}
\usepackage{appendix}
\usepackage[margin=2cm]{geometry}
\biboptions{sort&compress}
\usepackage[italicdiff]{physics}
 
\newtheorem{manualtheoreminner}{Theorem}
\newenvironment{manualtheorem}[1]{%
	\IfBlankTF{#1}
	{\renewcommand{\themanualtheoreminner}{\unskip}}
	{\renewcommand\themanualtheoreminner{#1}}%
	\manualtheoreminner
}{\endmanualtheoreminner}

\newtheorem{manualpropositioninner}{Proposition}
\newenvironment{manualproposition}[1]{%
	\IfBlankTF{#1}
	{\renewcommand{\themanualpropositioninner}{\unskip}}
	{\renewcommand\themanualpropositioninner{#1}}%
	\manualpropositioninner
}{\endmanualpropositioninner}

\newtheorem{manuallemmainner}{Lemma}
\newenvironment{manuallemma}[1]{%
	\IfBlankTF{#1}
	{\renewcommand{\themanuallemmainner}{\unskip}}
	{\renewcommand\themanuallemmainner{#1}}%
	\manuallemmainner
}{\endmanuallemmainner}

\pgfplotsset{compat=1.18} 
\journal{arXiv}

\begin{document}
	
	\begin{frontmatter}
			
		\title{Host diversity and adaptive vector feeding preferences shape disease burden in vector-borne diseases}

		\author{Shravani Shetgaonkar}
		\author{Anupama Sharma \footnote{Corresponding author e-mail:\textit{anupamas@goa.bits-pilani.ac.in}}}
		
		\affiliation{organization={Department of Mathematics, Birla Institute of Technology and Science, Pilani, K K Birla Goa Campus, Zuarinagar, Sancoale, Goa 403726, India}}
		
		\begin{abstract}
Vector-borne diseases often involve multiple host species that differ in their ability to sustain transmission. At the same time, vector feeding preferences can change in response to host availability and disease-control interventions, potentially altering disease dynamics in unexpected ways. We develop a two-host vector-borne disease model that links host diversity, adaptive vector feeding preferences, and disease transmission. We demonstrate that the effect of host diversity on disease transmission is mediated by vector feeding behavior and cannot be inferred from host abundance alone. In particular, we identify a critical threshold, $R_{0c}$, that determines whether shifts in vector preference amplify or suppress disease burden in a focal host. This threshold marks a qualitative transition in system behavior and provides a basis for predicting epidemiological responses to changes in host composition and vector behavior. Using adaptive dynamics, we further show that vector populations may evolve toward either specialist or opportunistic feeding strategies depending on host encounter rates and trade-off strength. Finally, we demonstrate that host-targeted interventions can induce adaptive changes in vector feeding behavior that reduce prevalence in the protected host while potentially increasing overall infection burden. Our results highlight how evolutionary responses of vector population can generate unexpected epidemiological outcomes and should be considered when designing disease-control strategies.		
		\end{abstract}
	\end{frontmatter}

\section{Introduction}\label{sec1}

Many vector-borne diseases circulate among multiple host species that differ in abundance, susceptibility, and transmission competence~\cite{martinez2021differential,medeiros2021mathematical}. In such systems, disease transmission depends not only on the epidemiological characteristics of individual hosts but also on how vectors distribute their bites among available host species. Consequently, vector feeding preferences can play a critical role in determining transmission pathways, reservoir importance, and disease burden across the host community~\cite{fairbanks2025quantifying}. Understanding these interactions is particularly important because host diversity can either amplify or dilute disease transmission~\cite{Miller}, yet the conditions leading to these contrasting outcomes remain poorly understood. Moreover, vector feeding preferences can shift in response to host availability and host defenses, suggesting that amplification and dilution effects are not fixed properties of host communities but may change over ecological and evolutionary timescales. As a result, there is a growing need to understand how host diversity and vector feeding behavior interact to determine disease persistence, prevalence, and control outcomes in multi-host systems.

The importance of vector feeding behavior in multi-host systems has been increasingly recognized~\cite{Althouse}. Since vectors do not distribute their bites uniformly among available hosts, differences in host preference can alter transmission pathways and the contribution of a particular host species to disease persistence. Moreover, when vectors preferentially feed on hosts that differ in transmission competence, changes in host community composition can produce markedly different epidemiological outcomes. For example, disease may persist when vectors exhibit a strong preference for a highly competent host, even when that host is relatively scarce compared with alternative hosts \cite{sulaimon2025potential}. Also, increasing the abundance of incompetent hosts has been shown to reduce transmission of Chagas disease by diverting vector bites away from more competent hosts \cite{zahid2020decoys}. These observations suggest that the epidemiological consequences of host diversity depend not only on host abundance and competence, but also on how vectors allocate their feeding effort among available hosts. Nevertheless, most mathematical models of vector-borne diseases have focused on single-host single-vector systems \cite{ross1911prevention,Macdonald,chitnisR_0,chamchod2011analysis,rock2015age,anwar2024mathematical}, limiting our understanding of how host diversity and vector host choice jointly shape disease dynamics.

Establishing a unifying principle of how host diversity influences VBD transmission is challenging, as its effect on disease dynamics is largely shaped by the ecological specifics of each system~\cite{yakob2010modelling, simpson, rivera2020relation,bilal2020complexity,chen2022host,wu2023spatial}. For example, whether host diversity amplifies, dilutes, or has no effect on disease prevalence depends on the interplay between vector preference and host transmission abilities, with amplification occurring when both host species are present, leading to maximal disease risk~\cite{Miller}. Moreover, a ``no-choice'' and ``two-choice'' experiment of feeding between humans, who were the preferred host, and sheep revealed the evolution of feeding preference of malaria vectors to be dependent on host availability \cite{bouafou2024host}. Hence, adaptive shifts in vector feeding preferences may become an important bottleneck to the long-term success of disease-control interventions. When vectors initially bite only competent hosts, time-dependent feeding preferences can increase disease prevalence and shift the timing of epidemic peaks~\cite{marini2017exploring}. Hence, amplification and dilution effects should not be viewed as static properties of a host community but as outcomes that may change as vector populations adapt their feeding strategies. Despite this, the evolutionary consequences of host diversity for vector host choice, and the resulting feedbacks on disease transmission, remain poorly understood.

In this study, we investigate how host diversity influences both disease transmission and the evolution of vector feeding preferences. We develop a two-host vector-borne disease model and use adaptive dynamics~\cite{geritz1997dynamics} to examine how ecological variation in host communities shapes vector host choice. { We identify conditions under which changes in vector preference reverse the epidemiological effect of host diversity, causing shifts between amplification and dilution regimes. We show that host encounter patterns can drive the evolution of either strong specialization on a preferred host or stable opportunistic feeding strategies. We further show that interventions that reduce contacts with a preferred host can redirect vector feeding toward alternative hosts, reducing prevalence in the protected host while potentially increasing overall infection levels.} Our results suggest that host diversity mediated amplification and dilution of disease burden is not fixed property of host communities, but may change as vector populations adapt their feeding behavior.

The paper is structured as follows: In Section \ref{sec2}, we formulate a compartmental model for VBD with two hosts, one vector, and vector preference. Section \ref{sec3} covers the computation of $R_0,$ and the analysis of equilibria and their stability. Section \ref{sec4} provides a sensitivity analysis of $R_0$. Section \ref{adapdyn} explores the evolutionary dynamics of vector feeding preferences using adaptive dynamics, and Section \ref{sec-ecoepi} couples the adaptive dynamics to epidemiological outcomes. Section \ref{sec-con} concludes with a discussion of the results.

\section{The model} \label{sec2}
We develop a model to study the spread of VBD in a population comprising two hosts ($h1$) and ($h2$) and a single vector ($v$).  Let $N_{h1}$ be the total population size of $h1$, which we assume to be constant. At time $t$  we divide $N_{h1}$ into three classes viz., susceptible $S_{h1}(t)$, Infected $I_{h1}(t)$ and Recovered $R_{h1}(t)$. Let the total population of $h2$ be constant and denoted as $N_{h2}$. At time $t$ we divide $N_{h2}$ into two compartments, viz., Susceptible $S_{h2}(t)$ and Infected $I_{h2}(t)$. We denote the total vector population as $M$. The vectors get infected by biting an infected host, and at time $t$, the vector population is divided into two compartments: Susceptible $S_v(t)$ and Infected $I_v(t)$.  A central feature of our model is that we incorporate {\em host preference,} i.e., vector exhibit a feeding preference for $h1$ over the other host $h2$. This plays an important role in transmission dynamics, and we model it explicitly by comparing the ratio of bites on $h1$ and $h2,$ normalised by their respective population size \cite{Miller}. If $\alpha_v>1$ (or $0<\alpha_v<1$) then the vector has a propensity for feeding on host species $h1$ (or $h2$). The vector has no preference when $\alpha=1$.
	
We consider the $SI$ model for the vector population, as the vector lifespan is short and they die before recovering from the disease. Moreover, host $h1$ is assumed to acquire temporary immunity following recovery, leading to an $SIRS$ model, whereas host $h2$ becomes immediately susceptible after recovery, leading to an $SIS$ model. The infected $h1$ population recovers at a per capita recovery rate $\mu_1$. After a period of $1/\delta_1$, they lose their immunity to the disease and become susceptible again. The infected $h2$ individuals recover and become susceptible again at a rate of $\mu_2$. The nonlinear ordinary differential equations governing the dynamics of the spread of the disease are as follows: 
	\begin{align}\label{1}
		\begin{split}
			\text{host 1 }
			&\begin{cases}
				\vspace{5px}\diff{S_{h1}}{t}= b_{1} N_{h1}-\frac{\alpha_{v}c \beta_{hv} I_{v}  S_{h1}}{\alpha_{v} N_{h1}+N_{h2}} + \delta_{1} R_{h1}-d_{1} S_{h1},\\\vspace{5px}
				\diff{I_{h1}}{t}=\frac{\alpha_{v} c \beta_{hv} I_{v} S_{h1}}{\alpha_{v} N_{h1}+N_{h2}}-(\mu_{1}+d_{1}) I_{h1},\\
				\diff{R_{h1}}{t}=\mu_{1} I_{h1}-(\delta_{1}+d_{1})  R_{h1},\\
			\end{cases}\\
			\text{host 2 }&\begin{cases}
				\vspace{5px}\diff{S_{h2}}{t}=b_{2} N_{h2}-\frac{c \beta_{hv}  I_{v}
					S_{h2}}{\alpha_{v} N_{h1}+N_{h2}} + \mu_2 I_{h2}-d_{2} S_{h2},\\\vspace{5px}
				\diff{I_{h2}}{t}=\frac{c \beta_{hv}  I_{v}  S_{h2}}{\alpha_{v} N_{h1}+N_{h2}}-(\mu_2+d_{2} ) I_{h2},\\
			\end{cases}\\
			\text{vector }&\begin{cases}
				\vspace{5px}\diff{S_{v}}{t}=(b_3-b_{31} M) M-\frac{c \beta_{vh}  (\alpha_{v} I_{h1}+I_{h2}) S_{v}}{\alpha_{v} N_{h1}+N_{h2}}-d_3 S_{v},\\\vspace{5px}
				\diff{I_{v}}{t}=\frac{c \beta_{vh} (\alpha_{v} I_{h1}+I_{h2}) S_{v}}{\alpha_{v} N_{h1}+N_{h2}}-d_3 I_{v},
			\end{cases}
		\end{split}
	\end{align}
	\noindent with initial conditions $S_{h1}(0)>0,$ $I_{h1}(0)>0,$ $R_{h1}(0)\ge0,$ $S_{h2}(0)>0,$ $I_{h2}(0)\ge0,$ $S_{v}(0)>0,$ $I_{v}(0)\ge0.$ \\
	
Here, we assume equal birth and death rates of the host species, denoted by $d_i (=b_i)$ for host population $hi$, $i=1,2$. This keeps the total host populations constant. Further, $b_3$ is the birth rate of the vector population, $b_{31}$ accounts for the crowding effect on vector births. The mortality rate of vectors is $d_3$, and $c$ is the biting rate of vectors on hosts per unit time. The transmission probabilities from infected hosts to susceptible vectors and from infected vectors to susceptible hosts are denoted by $\beta_{vh}$ and $\beta_{hv}$, respectively. All parameters are assumed to be positive.
To maintain analytical tractability while focusing on the effects of feeding preference on transmission dynamics in a two-host system, disease-induced mortality is not incorporated into the model.  The model parameters are summarized in Table~\ref{table1}.
	
	\begin{table}[ht!]
		\centering
		\begin{tabular}{|c|c|c|c|}
			\hline
			Parameter &  Description &Value & Source\\
			\hline
			$\beta_{hv}$  & transmission probability from vector to host & 0.5 & \cite{barbosa2018modelling, ruan2008delayed, clancy2024extinction} \\
			$\beta_{vh}$ & transmission probability from host to vector& 0.5 & \cite{kamiya2017epidemiological, ruan2008delayed} \\
			$ c$& biting rate &1 $\text{day}^{-1}$ & \cite{moghadas2017asymptomatic} \\
			$\alpha_v$ & feeding preference & varied & \\ 
			$\mu_1$ & recovery rate of $h1$& 0.5 $\text{day}^{-1}$ & \cite{laperriere2011simulation}\\ 
			$\delta_1$ & rate of waning immunity of $h1$ & 0.5 $\text{day}^{-1}$ & assumed \\ 
			$\mu_2$ & recovery rate of $h2$& 0.5 $\text{day}^{-1}$ & \cite{sulaimon2025potential}  \\ 
			$b_3$ & birth rate of vector& 5 $\text{day}^{-1}$ & \cite{multerer2019modeling} \\ 
			$d_3$ & death rate of vector& 0.8 $\text{day}^{-1}$ & \cite{medeiros2021mathematical}  \\ 
			$d_1$ & birth/death rate of $h1$&0.5 $\text{day}^{-1}$ & assumed \\ 
			$d_2$ & birth/death rate of $h2$ &0.5 $\text{day}^{-1}$ & assumed\\ 
			$N_{h1}$ & total population of $h1$ & 3000 & assumed\\ 
			$N_{h2}$ & total population of $h2$& 1000 & \cite{sulaimon2025potential} \\ 
			$K$ &carrying capacity of vector population& 15000 & (5 vectors per $h1$)\cite{clancy2024extinction} \\ 
			\hline
		\end{tabular}
		\caption{Parameters description and the biologically plausible parameter values used in the numerical simulation.}
		\label{table1}
	\end{table}

\section{Model analysis} \label{sec3}	
\noindent The total host population of $h1$ and $h2$ are constant as the model \eqref{1} satisfies 
    \[ \diff{N_{h1}}{t}=(d_1 - d_1)N_{h1}=0,\quad  \diff{N_{h2}}{t}=(d_2 - d_2) N_{h2}=0. \]
    By considering $S_{h1}(t)=N_{h1}-I_{h1}(t)-R_{h1}(t)$, $S_{h2}(t)=N_{h2}-I_{h2}(t)$ and $S_{v}(t)=M(t)-I_{v}(t)$ we reduce model \eqref{1}. Moreover, analyzing the model in terms of proportions of susceptible, infectious, and immune individuals simplifies the process. Therefore, we introduce a change of variables,
		$x_{1}=\frac{S_{h1}}{N_{h1}},\; y_{1}=\frac{I_{h1}}{N_{h1}},\; z_{1}=\frac{R_{h1}}{N_{h1}},\; x_{2}=\frac{S_{h2}}{N_{h2}},\; y_{2}=\frac{I_{h2}}{N_{h2}},\; x_{3}=\frac{S_{v}}{M},\; y_{3}=\frac{I_{v}}{M}$, so we have $x_1+y_1+z_1=1,\; x_2+y_2=1,\; x_3+y_3=1.$ Note that when $M=0$ we get $y_3=0$ as $I_v=0$.  The model can be written as follows:
			\begin{align} \label{req}
			\begin{split}
				\diff{y_{1}}{t}&=\frac{\alpha_{v}\beta_{hv}cM}{(\alpha_{v} N_{h1}+N_{h2})}\; y_3 (1-y_1-z_1)-(\mu_1+d_1)y_1,\\
				\diff{z_{1}}{t}&=\mu_{1} y_{1}-( \delta_{1}+d_{1}) z_{1},\\
				\diff{y_{2}}{t}&=\frac{\beta_{hv}cM}{(\alpha_{v} N_{h1}+N_{h2})}\; y_3  (1-y_2)-(\mu_{2}+d_{2}) y_2,\\
				\diff{y_3}{t}&=\frac{\beta_{vh}c}{(\alpha_{v} N_{h1}+N_{h2})}\;(\alpha_{v} N_{h1} y_1+N_{h2} y_2) (1-y_3)-d_3 y_3,\\
            \diff{M}{t}&=\big(b_3-d_3-b_{31}M \big) M,
			\end{split}
		\end{align}\\
with initial condition $y_{1}(0)>0,\;z_{1}(0)\ge 0, \;y_{2}(0)\ge0,\;y_{3}(0)\ge0$, $M(0)>0$. 
 We can verify the existence and uniqueness of the solution to the initial value problem \eqref{req} with the help of the existence and uniqueness theorem.

	To ensure well-posedness and exclude nonphysical negative populations, the model is required to admit positive and bounded solutions.

	\begin{manuallemma}{1}
		For nonnegative initial conditions, the solutions of \eqref{req} remain nonnegative, and the region $\Gamma$ is a positively invariant set for solutions of the system \eqref{req}, such that
		\small $$\Gamma=\Big\{(y_1,z_1,y_2,y_3,M)\in \mathbb{R}^{5}:0\le y_1+z_1<1,\;0\le y_2,\;y_3<1,\; 0\le M \le \frac{b_3-d_3}{b_{31}}=K \Big\}.$$ 
	\end{manuallemma}
\noindent The vector population exhibits a logistic growth rate with the carrying capacity $K>0$.
Now it suffices to focus on the model system \eqref{req} within the region $\Gamma$. 
	\subsection{Basic reproduction number $R_0$} 
	\noindent The basic reproduction number, denoted by $R_0,$ quantifies the number of secondary infectious cases generated by a single infected individual introduced within a fully susceptible population. We have computed $R_0$ for model system \eqref{req} using the Next Generation Matrix method \cite{diekmann2010construction}, as
	\begin{equation}\label{Ro}
		R_0=\frac{\beta_{hv}\beta_{vh}c^2 K}{(\alpha_v N_{h1}+N_{h2})^2d_3}\left(\frac{\alpha_{v}^{2} N_{h1}}{(\mu_{1}+d_1)} +\frac{N_{h2}}{(\mu_2+d_{2})}\right).
	\end{equation}
	For details, see \ref{AppA}.\\
	
	For two host systems, $R_0$  is the sum of two components, corresponding to the contributions from hosts $h1$ and $h2$. Each component constitutes the product of the total number of infections in vectors caused by one infected host and the number of infections in hosts arising from an infected vector. If there is a feeding preference of $\alpha_v$ for $h1$ and there are $M$ number of vectors then the total $h1$ population experiences $ \frac{ K c \alpha_v N_{h1}}{\alpha_v N_{h1} + N_{h2}}$ number of bites from vectors, as initially, the entire population is assumed to be susceptible. 
	Out of these only a fraction of $\frac{\beta_{hv}}{d_3} \frac{K c \alpha_v N_{h1}}{(\alpha_v N_{h1} + N_{h2})}$ bites by an infected vector potentially cause infection in $h1$. Similarly a fraction of $\frac{\beta_{hv}}{d_3} \frac{K c N_{h2}}{(\alpha_v N_{h1} + N_{h2})}$ bites by an infected vector can cause infection in $h2$. Furthermore, $\frac{\beta_{vh}}{(\mu_{1}+d_{1})} \frac{c \alpha_v}{(\alpha_v N_{h1} + N_{h2})}$ and $\frac{\beta_{vh}}{(\mu_2+d_{2})} \frac{c}{(\alpha_v N_{h1} + N_{h2})}$ fraction of bites on a infected $h1$ and $h2$ lead to mature infections in a vector. The above factors imply that,
	\begin{align*} 
		\begin{split} 
			R_0 &=\underbrace{\frac{\beta_{vh}\beta_{hv} K  c^{2} \alpha_{v}^{2}  N_{h1}}{(\alpha_{v} N_{h1}+N_{h2})^{2} d_3(\mu_{1}+d_{1})}}_ { \textstyle R_0 \text{ for } h1}  +
			\underbrace{{\frac{\beta_{vh}\beta_{hv}  K c^{2} N_{h2}}{(\alpha_{v} N_{h1}+N_{h2})^{2} d_3 (\mu_2+d_{2})}}}_{ \textstyle R_0 \text{ for } h2}.  
		\end{split}
	\end{align*} 
	Note that the extra squared term ($\alpha_{v}^{2}$) in the first expression accounts for the assumption that the vector prefers $h1$ with a preference rate of $\alpha_v$ for its feed both times while picking up and transferring the infection by biting at rate $c$. 

	\subsection{Disease free equilibrium}
\noindent At the disease-free equilibrium (DFE), all individuals are in the susceptible state, and there are no  infectious individuals in the population. The model system \eqref{req} always exhibit two disease-free equilibria (DFEs), denoted by $E_{00}=(0,0,0,0,0)$ and $E_{0K}=(0,0,0,0,K)$. 
The equilibrium $E_{00}$ indicates a VBD-free system in the absence of vector population. The equilibrium $E_{0K}$ corresponds to a disease-free state in which the vector population is at its carrying capacity $K$.
Now, to study the local stability analysis of DFE, we establish the following theorem:

\begin{manualtheorem}{1}\label{DFE_localstab}
The system \eqref{req} admits the following disease-free equilibrium
\begin{itemize}
\item[(i)] $E_{00}$ is unstable.
    \item[(ii)] $E_{0K}$ is locally asymptotically stable if $R_0<1$.
\end{itemize}

\end{manualtheorem}
\noindent See \ref{proof local stable DFE} for proof of Theorem \ref{DFE_localstab}.

\subsection{Endemic equilibrium} 
When $R_0>1$, the system \eqref{req} exhibits a unique endemic equilibrium,
\(
E^{*}=(y_1^{*},z_1^{*},y_2^{*},y_3^{*},K),
\)
where $K=\frac{b_3-d_3}{b_{31}}$ is the carrying capacity, and
$y_1^{*}$, $z_1^{*}$, and $y_2^{*}$ are expressed in terms of $y_3^{*}$ as follows:
\begin{equation*}
	\begin{aligned}
		& y_1^{*}=\frac{\alpha_v \beta_{hv}c K  y_3^{*}}{(m_3 y_3^{*} +m_1) (\alpha_{v} N_{h1}+N_{h2})},\quad z_1 ^{*}=\frac{\mu_1 y_1^{*}}{\delta_1+d_1}, \; \\ & y_2^{*} =\frac{\beta_{hv}c K y_3^{*}} {\beta_{hv}c K  y_3^{*} + (\alpha_{v} N_{h1}+N_{h2}) m_2}.
	\end{aligned}
\end{equation*}
Here
$ m_1 = \mu_1+d_1, \; m_2 = \mu_2+d_2,\; m_3 =  \frac{\alpha_{v}  \beta_{hv}cK}{(\alpha_{v} N_{h1}+N_{h2})}  (1 + \frac{\mu_1 } {\delta_1 + d_1 })$ and  $y_3^{*}$ is the positive solution of equation 
\begin{equation}\label{4}
	A y_3 ^{2}+ B y_3+C=0,
\end{equation}

\begin{align*}
	A&=\dfrac{\beta_{hv}\beta_{vh}c^2 K}{(\alpha_v N_{h1}+N_{h2})^2} \left( \frac{ \alpha_{v}^{2} N_{h1} \beta_{hv}c K}{(\alpha_{v} N_{h1}+N_{h2})} +N_{h2} m_3\right) + \frac{\beta_{hv}c K d_3 m_3}{(\alpha_{v} N_{h1}+N_{h2})},\\
	B&=\frac{\beta_{hv}c K d_3 m_1}{(\alpha_{v} N_{h1}+N_{h2})} 
	\left(1-\frac{\beta_{hv}\beta_{vh}c^2 K}{(\alpha_v N_{h1}+N_{h2})^2 d_3} \frac{\alpha_v^2 N_{h1}}{m_1} \right) \\&+ d_3 m_3 m_2 \left(1-\frac{\beta_{hv}\beta_{vh}c^2 K}{(\alpha_v N_{h1}+N_{h2})^2 d_3} \frac{N_{h2}}{m_2} \right)  + \frac{\beta_{hv}\beta_{vh}c^2 K}{(\alpha_v N_{h1}+N_{h2})^2} (\alpha_{v}^{2} N_{h1} m_2+N_{h2} m_1),\\
	C&=d_3 m_1 m_2-\frac{\beta_{hv}\beta_{vh}c^2 K}{(\alpha_v N_{h1}+N_{h2})^2} (\alpha_{v}^{2} N_{h1} m_2+N_{h2} m_1)=d_3 m_1 m_2 (1-R_0).
\end{align*}

\noindent Here $A>0$ and $C<0$ when $R_0>1$, hence \eqref{4} will have a unique positive solution for $y_3$ say $y_3^*$.
When $R_0\le1$, $A, B>0$, and $C\ge 0$, Descartes' rule of signs confirms that equation \eqref{4} has no positive real roots. Hence, there exists a unique endemic equilibrium point for model \eqref{req}.\\

For the stability analysis of the endemic equilibrium of the model system \eqref{req}, we employ Liapunov's second method \cite{Stephen} and obtain the following results.

\begin{manualtheorem}{2} \label{THM local}
	When $R_0>1$ the endemic equilibrium $E^{*}$ exists uniquely and is locally asymptotically stable within the region $\Gamma$ if the following conditions hold
	\begin{equation}\label{5}
		\begin{split}
			\frac{2 \beta_{hv}\beta_{vh}c^2 K \alpha_{v}^{2} N_{h1}} {(\alpha_v N_{h1}+N_{h2})^2}&<d_3(\mu_1+d_1),\\
			N_{h1}N_{h2}\left(\frac{2 \beta_{hv}\beta_{vh}c^2 K \alpha_{v}} {(\alpha_v N_{h1}+N_{h2})^2}\right)^2&<d_3^2(\mu_1+d_1)(\mu_2+d_2).
		\end{split}
	\end{equation}
\end{manualtheorem}

\begin{manualtheorem}{3}\label{THM global}
	The endemic equilibrium $E^{*}$ is globally asymptotically stable  in $\Gamma$ if the following conditions hold
	\begin{equation}\label{6}
		\begin{split}
			\frac{2 \beta_{hv}\beta_{vh}c^2 K \alpha_{v}^{2} N_{h1}} {(\alpha_v N_{h1}+N_{h2})^2}&<d_3(\mu_1+d_1),\\
			N_{h1}N_{h2}\left(\frac{2 \beta_{hv}\beta_{vh}c^2 K \alpha_{v}} {(\alpha_v N_{h1}+N_{h2})^2}\right)^2&<d_3^2(\mu_1+d_1)(\mu_2+d_2).
		\end{split}
	\end{equation}
\end{manualtheorem}

\noindent See \ref{proof local stable} and \ref{proof global stable} for proof of Theorem \ref{THM local} and Theorem \ref{THM global}, respectively.

\section{Parameter sensitivity of $R_0$} \label{sec4}
\noindent From the disease-control viewpoint, the key to designing effective interventions is to identify the parameters with the greatest impact on $R_0$. With this goal, we analyse the expression of $R_0$, and find that $R_0$ increases with the parameters $\beta_{hv}$, $\beta_{vh}$, $c$, and $K$ $\big( \frac{\partial R_0}{\partial \beta_{hv}}, \frac{\partial R_0}{\partial \beta_{vh}}, \frac{\partial R_0}{\partial c}, \frac{\partial R_0}{\partial K} > 0 \big)$, and decreases with increasing values of $d_1$, $d_2$, $d_3$, $\mu_1$, and $\mu_2$ $\big( \frac{\partial R_0}{\partial d_1} , \frac{\partial R_0}{\partial d_2}, \frac{\partial R_0}{\partial d_3}, \frac{\partial R_0}{\partial \mu_1}, \frac{\partial R_0}{\partial \mu_2} < 0 \big)$. We now analyze the parameters whose influence on $R_0$ depends on the values of other model parameters.

\subsection{Effect of vector feeding preference $\alpha_v$ on $R_0$}
\noindent To find how $R_0$ varies with respect to $\alpha_v$ we partially differentiate $R_0$ with respect to $\alpha_v$ while keeping all other variables constant,
\begin{equation} \label{R0_av}
	\frac{\partial R_0}{\partial \alpha_v}=\frac{2\beta_{hv}\beta_{vh}c^2 K N_{h1}N_{h2}}{(\alpha_{v} N_{h1}+N_{h2})^3d_3}\left(\frac{\alpha_v}{(\mu_{1}+d_1 )}-\frac{1}{(\mu_2+d_{2})}\right).
\end{equation}
\noindent We define the critical preference threshold
\[
\alpha_{vc}=\frac{\mu_1+d_1}{\mu_2+d_2},
\]
which is the ratio of the infectious removal rates of the two host species. This implies that $\alpha_{vc}$ compares the rates at which infected individuals leave the infectious class in the two hosts and is the inverse of the ratio of their mean infectious periods. When the vector preference satisfies $\alpha_v=\alpha_{vc}$, the basic reproduction number attains its minimum value,
\[
R_{0c}=\frac{\beta_{hv}\beta_{vh}c^2K}{\left((\mu_1+d_1)N_{h1}+(\mu_2+d_2)N_{h2}\right)d_3}.
\]
Furthermore, from~\eqref{R0_av}, when vectors exhibit a strong preference for host $h_1$ (i.e., $\alpha_v$ large), increasing the recovery rate of $h_1$ beyond the threshold $\mu_1>\alpha_v(\mu_2+d_2)-d_1$ reduces $R_0$. Conversely, when vectors predominantly prefer host $h_2$ $(\alpha_v\ll1)$, increasing the recovery rate of $h_2$ so that $\mu_2>(\mu_1+d_1)/\alpha_v-d_2$ reduces disease transmission. Thus, control efforts are most effective when they target the host species that is preferentially fed upon by the vector.
\\

\par
\begin{figure}[ht!]
	\centering
	\includegraphics[width=0.6\linewidth]{ 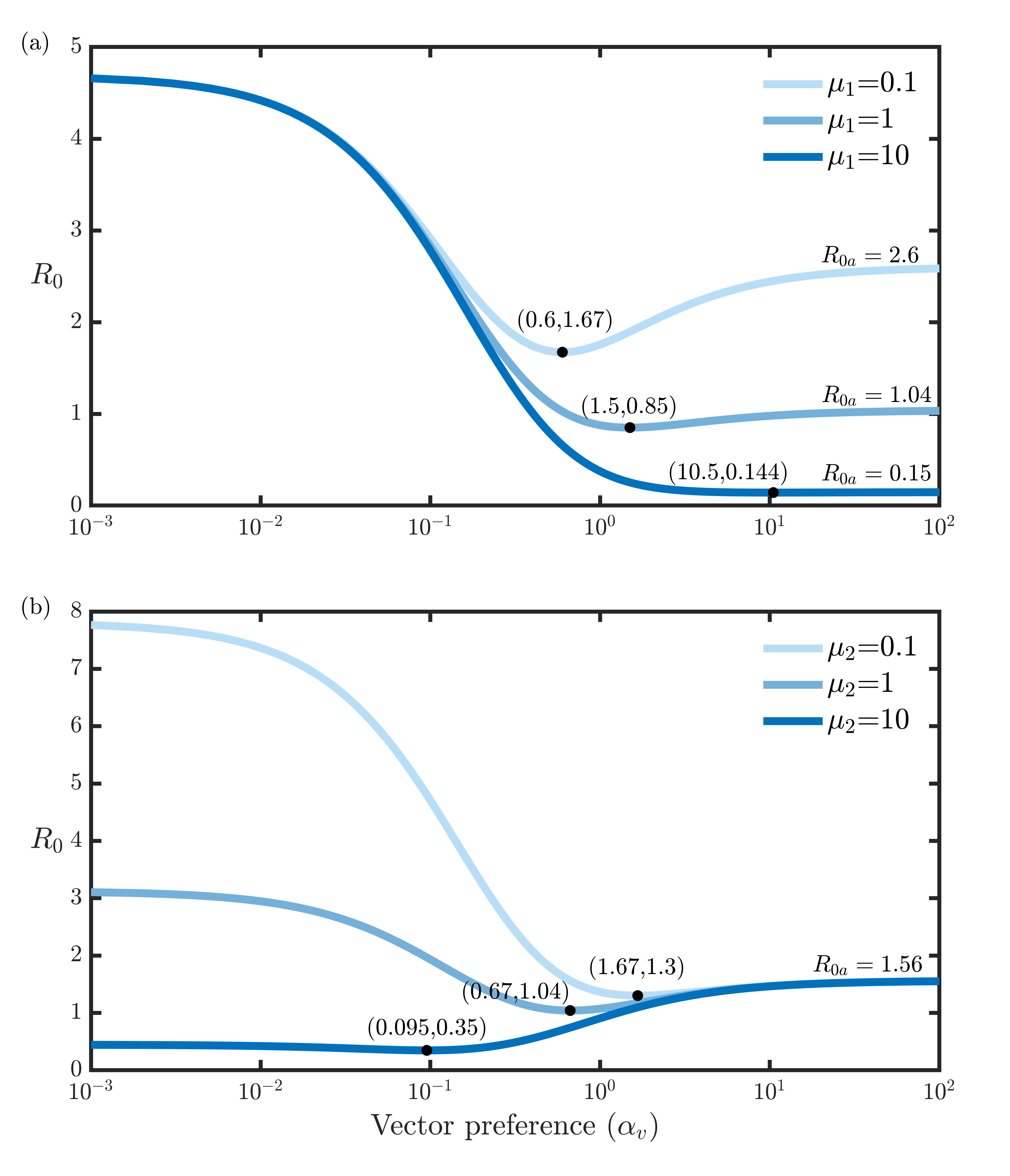}
	\caption{\textbf{Effect of recovery rate on the behavior of $\mathbf{R_0}$ as a function of vector preference.}  $R_0$ is varying with $\alpha_{v}$ for different values of $\mu_{1}$ in (a) and $\mu_2$ in (b). The black circle indicates the minimum value of $R_0$, which is attained at $\alpha_{vc}$ for all the curves. The value of $R_0$ converges to $R_{0a}$ with further increase in $\alpha_v$. $R_{0a}=$1.563 for all curves in (b), indicating its value is independent of $\mu_2$. The lowest value of $R_0$ is attained when the vector preference is large for the less competent host.}
	\label{fig2}
\end{figure}

 Figure \ref{fig2}(a)-(b) shows that increasing the recovery rate of the preferred host results in an asymptotically smaller value of $R_0$. From Figure \ref{fig2}(b), we see that changing the recovery rate of $h2$ does not affect the value at which $R_0$ converges. This is because $R_{0a}=\lim\limits_{\alpha_v \rightarrow \infty} R_0=\frac{\beta_{hv}\beta_{vh}c^2 K}{d_3(\mu_1+d_1)N_{h1}}$ does not depend on $\mu_2$. Hence, the spread of disease will be reduced if individuals of the preferred host class leave the infectious compartment at a faster rate than those not preferred. Note that $\alpha_{vc}$ is a measure of relative disease transmission of $h2$ as compared to $h1$. If $0<\alpha_{vc}<1$ or $\alpha_{vc}>1$ then the more competent host is $h1$ or $h2$ respectively. Hence, our analysis indicates that a decrease in the competency of the highly preferred host can result in a reduction of $R_0.$    

{ \subsection{Effect of preferred host population $N_{h1}$ and non-preferred host population $N_{h2}$ on $R_0$}

To understand how  $R_0$ varies with $N_{h1},$ we differentiate $R_0$ with respect to $N_{h1}$ by keeping other variables constant. We have
$$\frac{\partial R_0}{\partial N_{h1}}=\frac{\beta_{hv}\beta_{vh}c^2 K \alpha_{v}}{(\alpha_{v} N_{h1}+N_{h2})^3d_3}\left(\frac{\alpha_v N_{h2}-\alpha_{v}^2 N_{h1}}{(\mu_{1}+d_1 )}-\frac{2N_{h2}}{(\mu_2+d_{2})}\right).$$
From above we see that $R_0$ increases with $N_{h1}$ if
\begin{align}\label{ConR0Nh1}
\alpha_v > 2 \alpha_{vc} \text{ and }
N_{h1} < \frac{N_{h2}}{\alpha_v} \left(1 - \frac{2\alpha_{vc}}{\alpha_v}\right)=N_{h1c},
\end{align}
otherwise, $R_0$ decreases as $N_{h1}$ increases for $N_{h1}>N_{h1c}$. Conversely, if 
\(
\alpha_v < 2\alpha_{vc},
\)
$R_0$ decreases monotonically as $N_{h1}$ increases. Accordingly, $R_0$ reaches a global maximum at  $N_{h1} = N_{h1c}, \text{ for} \; \alpha_v > 2 \alpha_{vc}.$
This result implies that, as the population of $h1$ grows, increasing its recovery rate $\mu_1$ above  
\(
\mu_1 \ge \frac{\alpha_v (\mu_2 + d_2)}{2} - d_1
\)
is sufficient to mitigate the spread of the vector-borne disease.\\

Similarly, we compute the partial derivative of $R_0$ with respect to $N_{h2}$ as,
$$\frac{\partial R_0}{\partial N_{h2}}=\frac{\beta_{hv}\beta_{vh}c^2 K}{(\alpha_{v} N_{h1}+N_{h2})^3d_3}\left(\frac{\alpha_v N_{h1}-N_{h2}}{(\mu_2+d_{2})}-\frac{2\alpha_v^2 N_{h1}}{(\mu_{1}+d_1)}\right).$$
Analysis of this expression reveals that $R_0$ increases with $N_{h2}$ if 
\begin{align}\label{ConR0Nh2}
2 \alpha_v < \alpha_{vc}
\text{ and }
N_{h2} < \alpha_v N_{h1} \left(1 - \frac{2 \alpha_v }{\alpha_{vc}}\right)=N_{h2c},
\end{align}
and $R_0$ decreases as $N_{h2}$ increases for $N_{h2}>N_{h2c}$. 
Conversely, if 
\(
2 \alpha_v > \alpha_{vc},
\)
$R_0$ decreases monotonically as $N_{h2}$ increases. Consequently, $R_0$ attains a global maximum at  
$
N_{h2} =N_{h2c}, \text{ for } 2\alpha_v < \alpha_{vc}.
$ Therefore, as the population of $h2$ grows, increasing the recovery rate $\mu_2$ above  
\(
\mu_2 \ge \frac{\mu_1 + d_1}{2 \alpha_v} - d_2
\) 
is sufficient to reduce the spread of the vector-borne disease. This demonstrates that the effect of the host populations and their recovery rates on $R_0$ depends on the value of $\alpha_{vc}$. Specifically, $R_0$ increases with the size of $h1$ or $h2$ population when vector preference is greater than twice or less than half the value of $\alpha_{vc}$, respectively.\\

We next examine the dependence of $R_0$ on the vector-host interaction parameter $\alpha_v$, highlighting its critical threshold $\alpha_{vc}$ for a fixed host population. As shown in Figure~\ref{fig3}, $R_0$ attains its minimum at $\alpha_v=\alpha_{vc}$. Moreover, $R_0$ is minimum when vectors preferentially feed on the more abundant but less competent host. The green curves denote the threshold host population sizes, namely $N_{h1c}$ in Figure~\ref{fig3}(a) and $N_{h2c}$ in Figure~\ref{fig3}(b). Note that, for a fixed value of $\alpha_v$, $R_0$ initially increases as the size of the host population increases and attains a maximum value at the green curve. When the host population grows beyond this threshold, the value of $R_0$ decreases as the host abundance reduce the probability that vectors encounter an infected individual in the preferred host population. \\

\begin{figure}[ht!]
	\centering
	\includegraphics[width=0.8\linewidth]{ 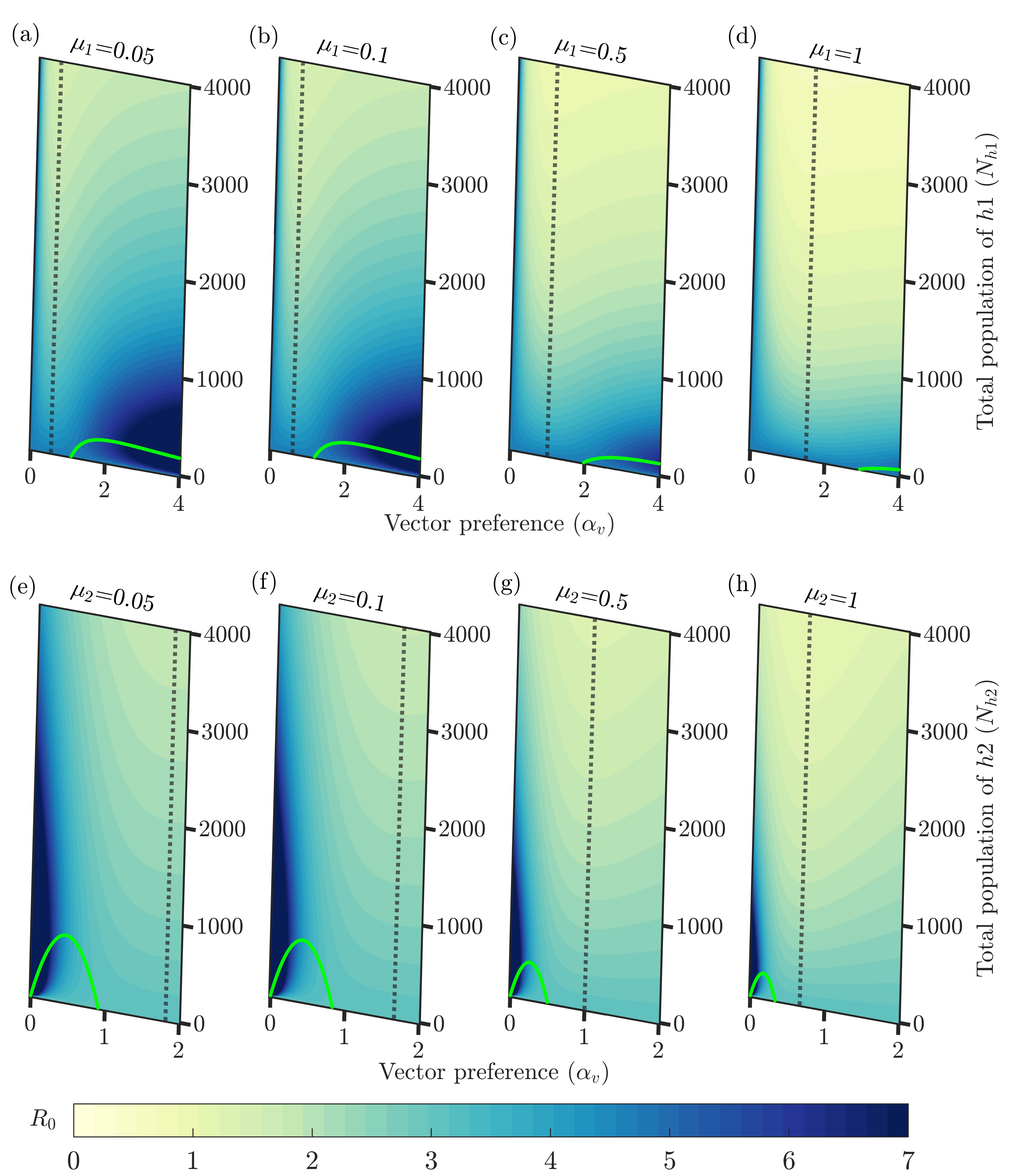}
	\caption{\textbf{Variation of $\mathbf{R_0}$ against changes in vector preference and size of host population for increasing recovery rates.} $R_0$ as a function of $\alpha_v$ is varying with: $N_{h1}$ for different values of $\mu_1$ in (a)-(d) and $N_{h2}$ for different values of $\mu_2$ in (e)-(h). Darker gradient colours correspond to higher values of $R_0$. Green curve represents the threshold value of the host population at which $R_0$ attains maximum. The highest value of $R_0$ is attained when the vector prefers the highly competent host. The dotted line denotes the value of $\alpha_{v}$ at $\alpha_{vc}$ for all panels. In (e)-(h), the parameter $M=30000$, the rest were as in Table \ref{table1}.	}
	\label{fig3}
\end{figure}

Figure \ref{fig10} presents the variance-based global sensitivity analysis of $R_0$ to the model parameters using Partial Rank Correlation Coefficient (PRCC). The parameter ranges considered in Figure \ref{fig10}(a) and \ref{fig10}(b) satisfy conditions \eqref{ConR0Nh1} and \eqref{ConR0Nh2} corresponding to regimes where $R_0$ increases with $N_{h1}$ and $N_{h2},$ respectively. The vector biting rate and death rate have the highest positive and negative PRCC indices, respectively. Further, $\alpha_v$ shows a positive correlation with $R_0$ in Figure \ref{fig10}(a) and a negative correlation in Figure \ref{fig10}(b), indicating the effect of feeding preference on disease spread. This is consistent with the trend observed in Figure \ref{fig3}, where $R_0$ increased with rising feeding preference for the preferred host. Moreover, the PRCC analysis also shows that recovery and death rates of the preferred host have a more pronounced negative impact on $R_0$ in comparison to the recovery and death rates of the non-preferred host. The transmission parameters $\beta_{vh}$ and $\beta_{hv}$ exhibit comparable PRCC values in both parameter regimes, suggesting that transmission in both the host-to-vector and vector-to-host directions contributes similarly to $R_0$. \\

\begin{figure}[ht!]
	\centering
	\includegraphics[width=0.8\linewidth]{ 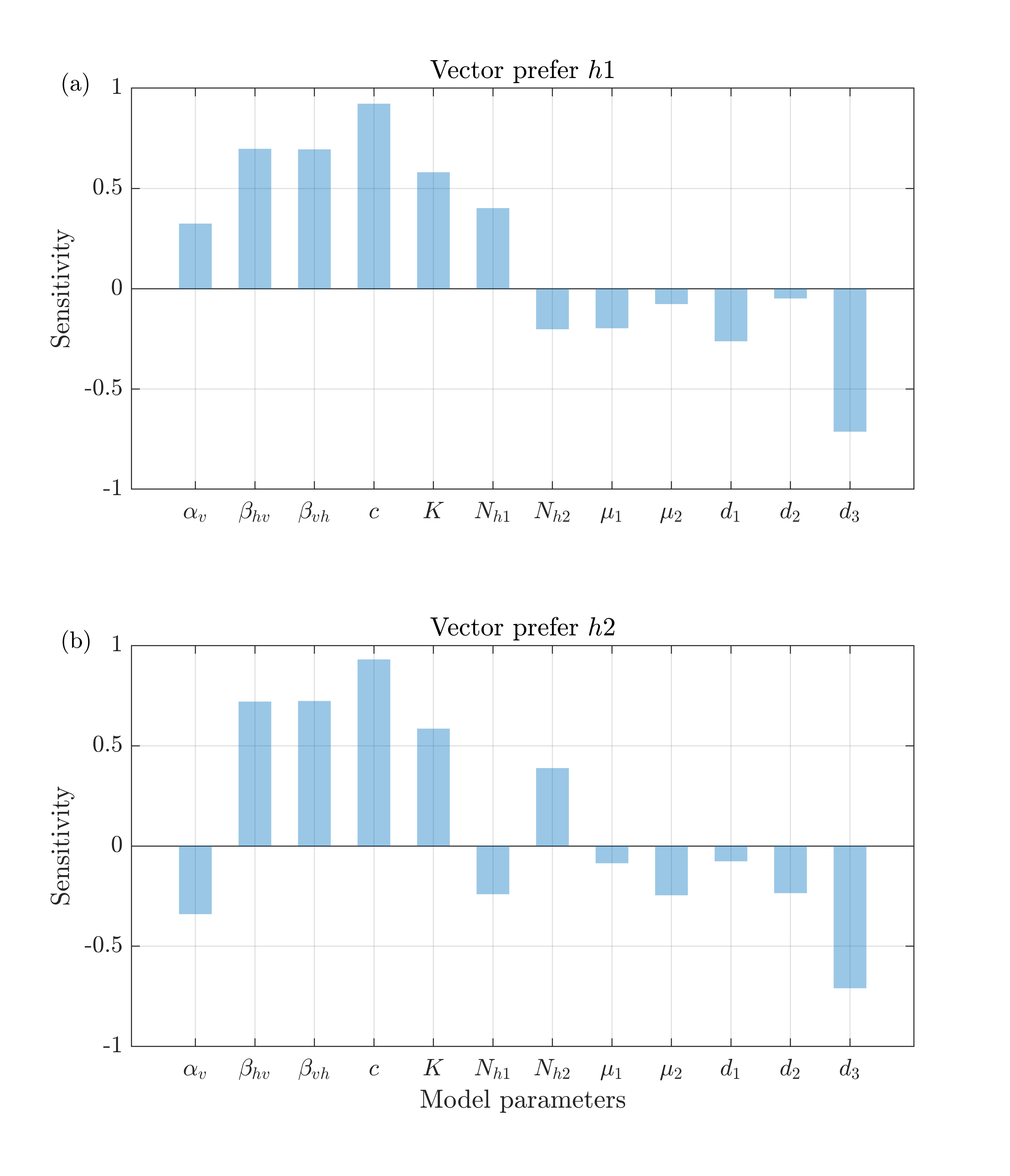}
	\caption{\textbf{Partial Rank Correlation Coefficient (PRCC) analysis of basic reproduction number $\mathbf{R_0}$, obtained using LHS with 1000 samples.} Positive PRCC value indicates that $R_0$ will increase when the parameter increases and vice versa. Sampling ranges used for panel (a) $\alpha_v=[3,7],\beta_{hv}=[0.1,1],\beta_{vh}=[0.1,1],c=[0.1,3],K=[7000,30000],N_{h1}=[0,500],N_{h2}=[5000,7000],\mu_{1}=[0.1,0.5], \mu_2=[0.5,0.1],d_1=[0.1,0.5],d_2=[0.5,0.1],d_3=[0.1,1]$ satisfy conditions \eqref{ConR0Nh1}, and for (b) $\alpha_v=[0.15,0.35],N_{h1}=[5000,7000],N_{h2}=[0,500],\mu_{1}=[0.5,1], \mu_2=[0.1,0.5],d_1=[0.5,1],d_2=[0.1,0.5]$ satisfy conditions \eqref{ConR0Nh2}.} 
	\label{fig10}
\end{figure}

\begin{figure}[ht!]
	\centering
	\includegraphics[width=0.8\linewidth]{ 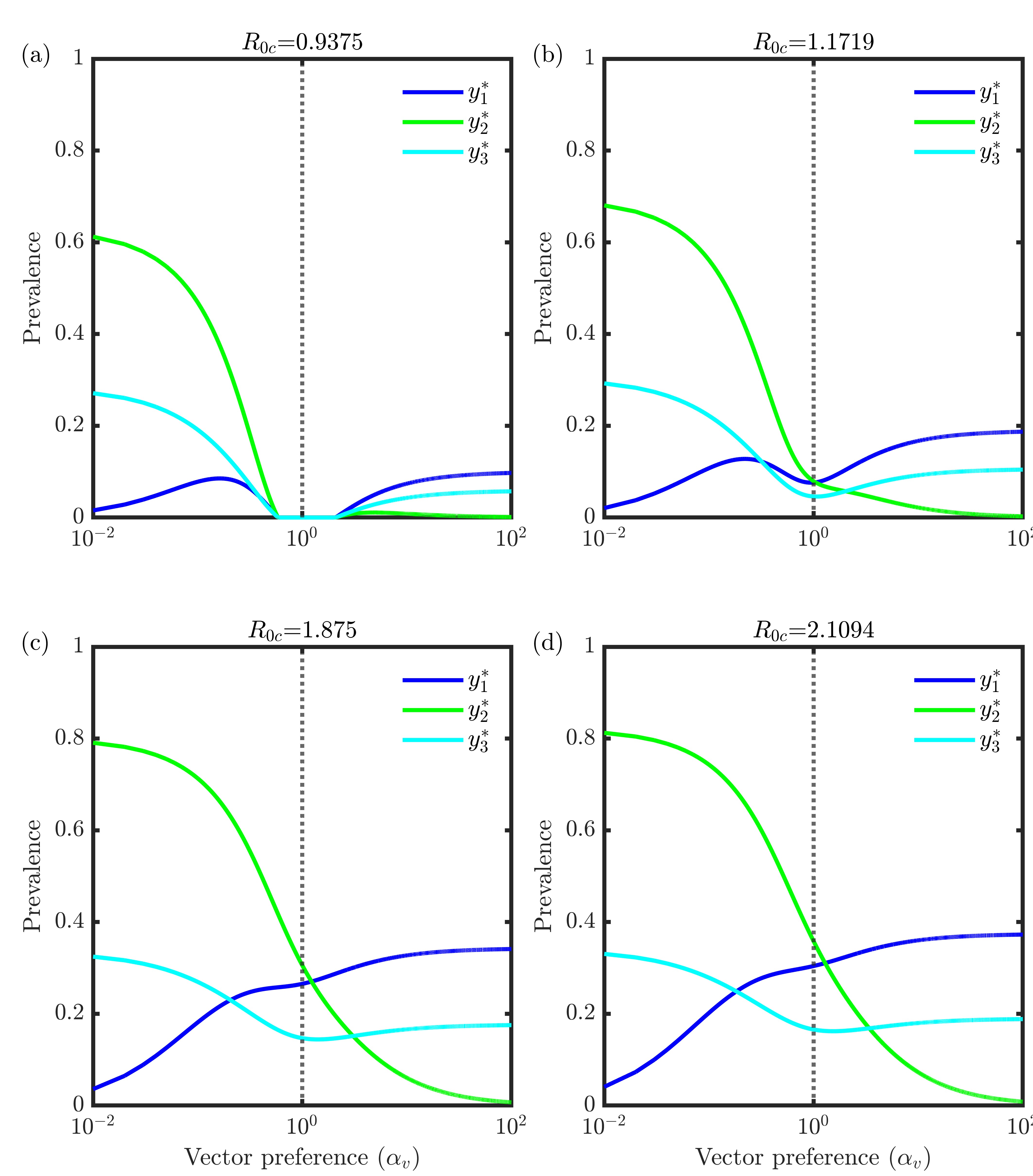}
	\caption{\textbf{The effect of $\mathbf{R_{0c}}$ with shifting vector preference on the behavior of equilibrium prevalence.} The transmission probability from vector to host is kept at (a) $\beta_{hv}=0.4$, (b) $\beta_{hv}=0.5$, (c) $\beta_{hv}=0.8$, (d) $\beta_{hv}=0.9$ to vary $R_{0c}$. The dashed line denotes the value of $\alpha_{v}$ at $\alpha_{vc}$. $R_{0c}$ is near 1 in (a)-(b) and $y_1^*$ has hump shaped behavior below $\alpha_{vc}$. In (c)-(d) $R_{0c}$ is higher and $y_1^*$ increases monotonically with $\alpha_v$. Other parameters were as in Table \ref{table1}.}
		\label{fig4}
\end{figure}

Figure \ref{fig4} shows the changes in the equilibrium disease prevalence for different values of $\alpha_v.$ It indicates that, depending on the value $R_{0c}$, increasing vector preference towards $h2$ can trigger a resurgence of VBD prevalence in $h1$, $y_1^*$. This implies that when $R_{0c}$ is close to unity, $y_1^{*}$ exhibits a {\em hump-shaped} behavior for $0<\alpha_v<\alpha_{vc}$, as shown in Figure  \ref{fig4}(a)-(b). In this regime, a gradual increase in  $\alpha_v$ initially raises the prevalence in $h1$, as vectors continue to feed substantially on $h2$, which serves as an alternative reservoir of infection, thereby enabling subsequent transmission to $h1$. As $\alpha_v$ approaches $\alpha_{vc}$, this trend reverses and the declining preference for $h2$ reduces the number of infected $h2$ individuals, lowering the infectious reservoir available to vectors and consequently reducing transmission to both hosts. In contrast, when $R_{0c}$ is sufficiently greater than one (owing to high values of \( \beta_{hv} \)), $y_1^{*}$ increases monotonically with increasing $\alpha_v$, whereas $y_2^{*}$ increases with decreasing $\alpha_v$, as shown in Figure \ref{fig4}(c)-(d). In this regime, transmission remains sufficiently strong that increasing vector preference toward $h1$ continually enhances transmission within the preferred host population while reducing transmission in $h2.$ Overall, these results suggest that adaptive shifts in vector preference can redistribute disease transmission between hosts and generate non-monotonic prevalence patterns in multi-host systems.

\section{Adaptive dynamics of vector host preference}\label{adapdyn}

In a single-host system, protecting the host can effectively reduce disease prevalence. However, this principle does not extend to multi-host systems such as ours, where community composition becomes a critical determinant of disease risk. When one host become ``functionally unavailable'' to vectors, for example through protective measures, an important evolutionary question is whether vectors evolve toward generalism or retain host specialization. Moreover, how will such evolutionary shifts affect the long-term dynamics of disease transmission?

To address these questions, we use adaptive dynamics to study the evolution of vector feeding preference and then examine how the resulting evolutionary outcomes influence disease transmission. We start by assuming that the dynamics of vector population depend on their innate feeding preferences and the strength of the trade-off related to mastering defense based on host type. Denote $\sigma_{h1}$ as the probability that the vector attacks $h1$ if encountered. We then assume that $\sigma_{h1}=1$ as the vectors have an inclination to feed on $h1$. The probability of attacking $h2$ for feeding if encountered is indicated by $\sigma_{h2}$ and is considered an evolving trait among vectors. Let $\mathcal{E}_{h1}$ and $\mathcal{E}_{h2}$ represent the encounter rates of $h1$ and $h2$ respectively. The probability that the vector survives till it detects a host is obtained by the following relation
$$p_1=\frac{\digamma p_{2}}{1-p_{2}(1-\digamma)},$$
Here $p_{2}$ and $\digamma$ are the probabilities of surviving and locating a host during a single host-seeking period, in which  $\digamma=1-e^{-(\sigma_{h1}\mathcal{E}_{h1}+\sigma_{h2}\mathcal{E}_{h2})}$ \cite{stone2018evolution}.

Also, while attempting to feed, some vectors may die fending off the host defense. Moreover, the vector's likelihood of evading the defensive behavior of a host enhances if it always attacks that specific host type. However, this could undermine its defense against the other host species. Let the mortality caused due to host defense be at maximum intensity of $\nu$, $0<\nu<1$. And the inverse of the trade-off strength be indicated by $\chi$. According to \cite{ravigne2009live}, such a trade-off can be described to obtain the probability of the vector surviving a feed on $h1$ and $h2$:
\begin{align*}
	p_{h1}&=1-\nu \varphi^\chi,\\
	p_{h2}&=1-\nu(1-\varphi)^\chi.
\end{align*}
Here $\varphi=\frac{\sigma_{h2}\mathcal{E}_{h2}}{\sigma_{h1} \mathcal{E}_{h1}+\sigma_{h2} \mathcal{E}_{h2}}$ is the conditional probability that $h2$ is fed upon after being located, and ($1-\varphi$) is the conditional probability of feeding on $h1$. Since $q_1=p_1 (\varphi p_{h2} +(1-\varphi) p_{h1})$ accounts for the vector's survival probability of foraging \cite{stone2018evolution}, $p_{rc}= q_1 q_2$ is the probability of surviving the whole cycle, where $q_2$ is the probability that the vector survives the phase after feeding on a host successfully. During this stage, it rests and reproduces before consuming another meal. Let $\tau_2$ be the duration of this second phase. To associate the daily survival rates of the vector population with their feeding intervals, we consider a similar approach as \cite{le2007elaborated}.
Assume that a normal foraging bout lasts for $\tau_f$ time. We can find the duration of one complete feeding cycle by accounting for the number of trials carried out to complete the cycle, $$ \tau_1=\frac{\tau_f}{1-p_{2}(1-\digamma)}.$$ 
The vector biting rate can be expressed as, $c=(\tau_1+\tau_2)^{-1}.$ Moreover, the death rate of the vector can be determined using $d_3=-\log(p_3)$ where $p_3=p_{rc}^c$ is the probability of surviving every day. 
Also, if $f_{rc}$ denotes the mean number of offspring produced per reproductive cycle, then the daily birth rate is $ b_3= f_{rc} c$. \\

To probe the influence of host encounter rate on the evolution of trait among vectors to attack $h2$ ($\sigma_{h2}$), we employ the adaptive dynamics approach. Consider the equilibrium value of the resident population and evaluate the invasion fitness of the mutant by finding its initial growth rate as
\begin{align*}
	\psi_{\sigma_{h2(r)}}(\sigma_{h2(m)})&=b_{3,\sigma_{h2(m)}}-d_{3,\sigma_{h2(m)}}-b_{31} M^{\star}, \text{ here }
	M^{\star}=\frac{b_{3,\sigma_{h2(r)}}-d_{3,\sigma_{h2(r)}}}{b_{31}}.
\end{align*}

The invading genotype will be evolutionarily advantageous if $\psi_{\sigma_{h2(r)}}(\sigma_{h2(m)})>0$. Here $\sigma_{h2(r)}$ and $\sigma_{h2(m)}$ denote the strategy used by the resident and mutant populations. At the evolutionary singular strategies (ESS), $\sigma_{h2}^*$, we have \[\left[\dfrac{\partial\psi_{\sigma_{h2(r)}}(\sigma_{h2(m)})}{\partial\sigma_{h2(m)}}\right]_{\sigma_{h2(r)}=\sigma_{h2(m)}^*,\atop \sigma_{h2(m)}=\sigma_{h2(m)}^*}=0.\]

Using the expression of invasion fitness and its derivative at the ESS, we can study possible evolutionary outcomes in the system. To understand the conditions on host community composition that lead the vector population to evolve no feeding preference or to adopt an opportunistic feeding strategy, we constructed pairwise invasibility plots (PIPs). Figure \ref{fig5:PIP}(a) shows that selection favors complete specialization on $h1$ in an environment favoring higher encounters with $h1$ than $h2$, that is, $\sigma_{h2}^*$ will approach 0 for all initial strategies. Whereas an ESS emerges when both the hosts are encountered at comparable rates or if $h2$ is encountered more frequently, as shown in Figure \ref{fig5:PIP}(b)-(d).

\begin{figure}[ht!]
	\centering
	\includegraphics[width=0.6\linewidth]{ 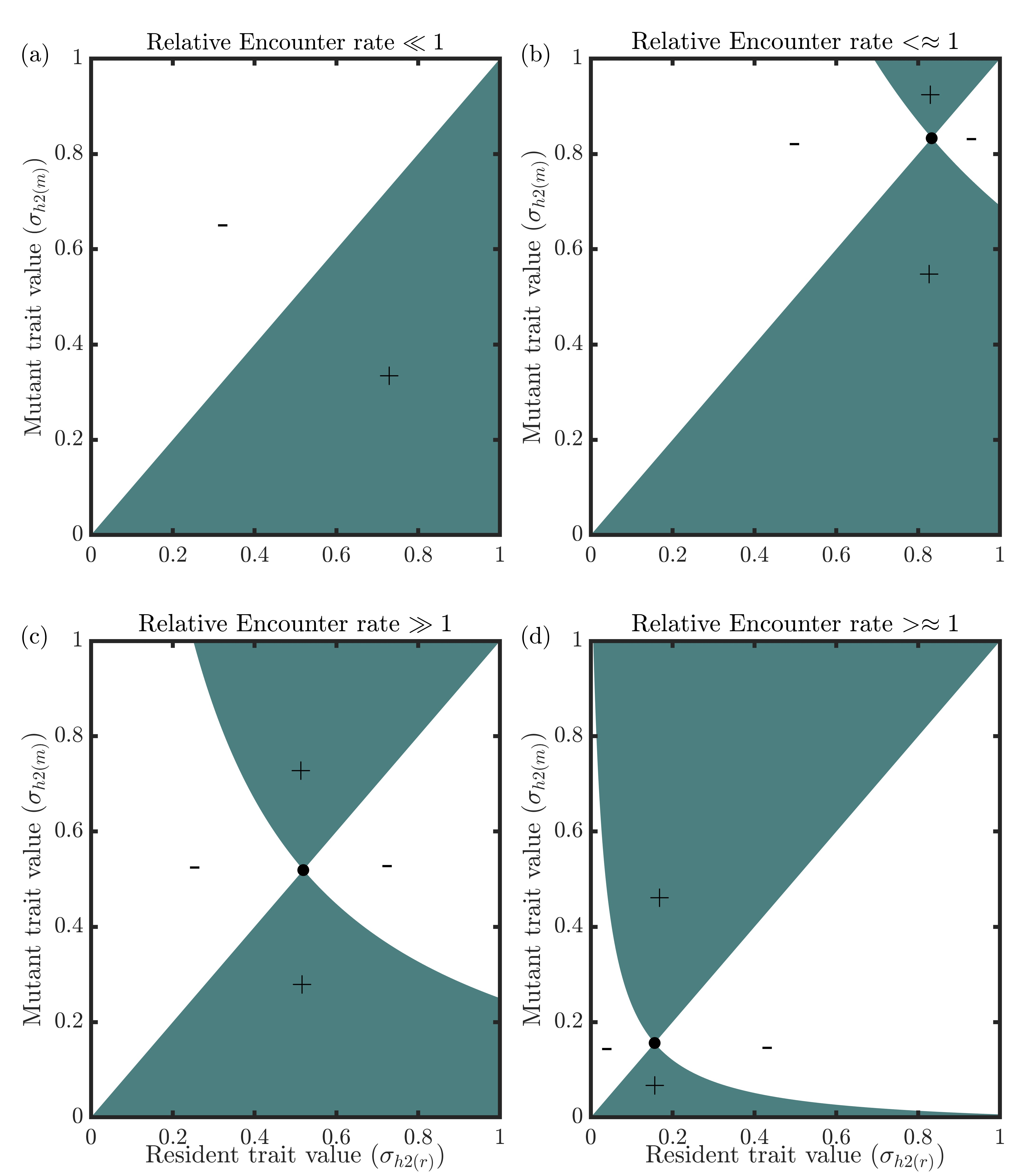}
	\caption{\textbf{Pairwise invasibility plots for vector trait to bite $\mathbf{h2}$ for relative encounter rate at  (a) $\mathbf{\frac{\mathcal{E}_{h2}}{\mathcal{E}_{h1}}=0.5}$, (b) $\mathbf{\frac{\mathcal{E}_{h2}}{\mathcal{E}_{h1}}=0.95}$, (c) $\mathbf{\frac{\mathcal{E}_{h2}}{\mathcal{E}_{h1}}=1.5}$, (d) $\mathbf{\frac{\mathcal{E}_{h2}}{\mathcal{E}_{h1}}=5}$.} The growth of the mutant strategy is indicated by the shaded region representing positive invasion fitness. The white area indicates negative invasion fitness. $h1$ is more abundant in (a)-(b) and $h2$ is more abundant in (c)-(d). Positive value for $\sigma_{h2}^*$ (black circle) is attained when the value of relative encounter rate is greater than or near to 1, otherwise $\sigma_{h2}^*=0$. Moreover, the evolutionary endpoint occurs at lower value when the encounter rate of $h2$ is very high (i.e. when $\frac{\mathcal{E}_{h2}}{\mathcal{E}_{h1}} \gg 1$) as compared to when they are nearly equal. Following from \cite{stone2018evolution}, the parameter values used in Figure \ref{fig5:PIP} are $b_{31}=0.0005,p_2=0.95$, $q_2=0.95$, $\nu=0.5,\; \chi=1$, $\tau_2$=2.5 day, $\tau_f=0.1$ day, $f_{rc}=5$.}
	\label{fig5:PIP}
\end{figure}
To understand how vector feeding behavior is evolutionarily linked to their ability to overcome host defenses, we calculate ESS for varied trade-off strengths. We found that $\sigma_{h2}^*$ vanishes as $\chi$ increases from 0 to a slightly high value, that is, strong to moderately weak trade-off strength, as shown in Figure \ref{fig6:Tradeoff}(a). This indicates that the vector's tendency is to specialize in feeding on $h1$ when encountered in greater numbers. Moreover, evolutionary endpoints attain values between 0 and 1 for further weakening of trade-off strength. In Figures \ref{fig6:Tradeoff}(b)-(d), the encounter rate with $h2$ is higher or similar to $h1$, and a nonzero ESS exists. These figures show that increasing the encounter rate of h2 expands the parameter space over which the trait of biting h2 is evolutionarily favored. We noted that for a weak trade-off (high value of $\chi$), the singular point shifts closer to 0. Hence, the value of $\sigma_{h2}^*$ is higher for similar host encounter rates or if the trade-off strength is moderate.

\begin{figure}[ht!]
	\centering
	\includegraphics[width=0.6\linewidth]{ 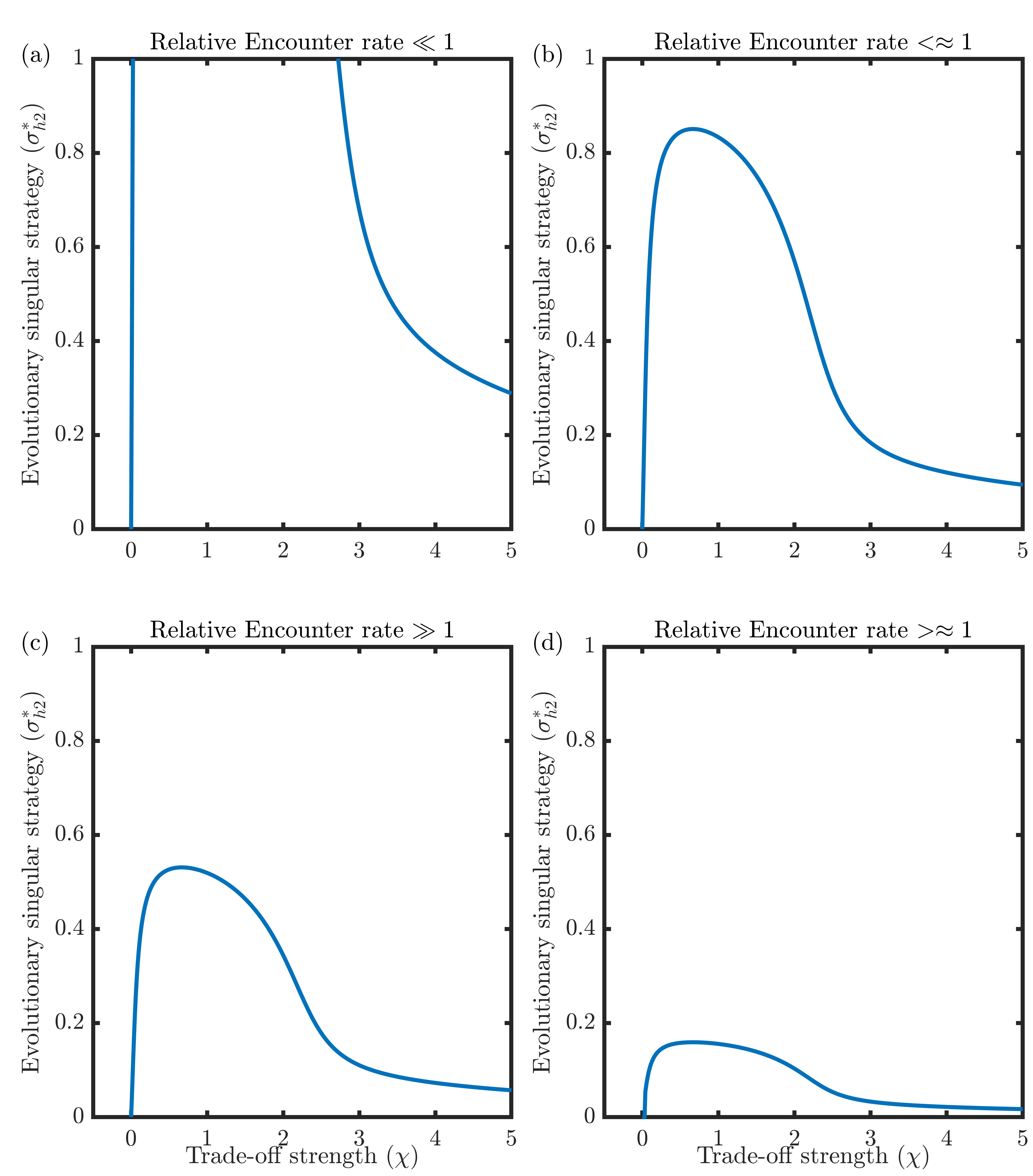}
	\caption{\textbf{Evolutionary 
			Singular strategies $\mathbf{\sigma_{h2}^{*}}$, obtained for different trade off strength, $\mathbf{(1/\chi)}$ for various level of encounter rates}. The trade-off between performance and preference is varied from strong ($\chi<1$) to weak ($\chi>1$). Note that in (a), the value of $\sigma_{h2}^{*}$ occurs between 0 and 1 for weak trade of strength values. In (b)-(d), the value of $\sigma_{h2}^{*}$ is high for intermediate trade-off values and decreases as trade-off strength becomes weak. The parameters were as in Figure \ref{fig5:PIP}.}
	\label{fig6:Tradeoff}
\end{figure}
To formalise the relationship between the evolutionary singular point and the trade-off strength, we present the following proposition:
\begin{manualproposition}{1}\label{singular_decrease}
	{\em Assume that $\mathcal{E}_{h1}=\mathcal{E}_{h2}=\mathcal{E}$ and $\chi>2$. Then $\sigma_{h2}^*$ decreases monotonically as $\chi$ increases if the following conditions are satisfied:
	\begin{align} \label{condition}
		\begin{split}
			(i)\quad&(\sigma_{h2}^{*})^{\chi -1} (\sigma_{h2}^{*}-\chi)+\chi \sigma_{h2}^{*}-1\le 0,\\
			(ii)\quad&(\sigma_{h2}^{*})^{\chi-1}+1-(\chi-1)(\sigma_{h2}^{*})^{\chi-2}\ge 0, \\
			(iii)\quad  &   \left(1-p_2 e^{-\mathcal{E}(1+\sigma_{h2}^*)}\right) \bigg(\frac{1}{(1+\sigma_{h2}^*)}+\frac{(\chi+1)(1+\sigma_{h2}^*)^\chi-\nu (1+\chi (\sigma_{h2}^{*})^{\chi-1})}{\left(1+\sigma_{h2}^{*}\right)^{\chi+1}-\nu\left(\sigma_{h2}^*+\left(\sigma_{h2}^{*}\right)^{\chi}\right) }\bigg)\ge2 p_2 \mathcal{E} e^{-\mathcal{E}(1+\sigma_{h2}^*)},\\
			(iv)\quad & \frac{\tau_f p_2 \left(\log(q_1 q_2)^{-1}-f_{rc}\right)}{\tau_f+\tau_2\left(1-p_2e^{-\mathcal{E}\left(1+\sigma_{h2}^{*}\right)}\right)} >\frac{(1-p_2)}{\left(1- e^{-\mathcal{E}\left(1+\sigma_{h2}^{*}\right)}\right)^2}.
		\end{split}
	\end{align}}
\end{manualproposition}
The proof of Proposition \ref{singular_decrease} is outlined in  \ref{pf:prop_5.1}. \\

Furthermore, we examined how the evolutionary singular point $\sigma_{h2}^{*}$ responds to changes in the trade-off strength. We found that when the trade-off is weak, i.e., when $\frac{1}{\chi} < \frac{1}{2}$ ($\chi > 2$), the value of $\sigma_{h2}^{*}$ decreases monotonically as the trade-off becomes even weaker ($\chi$ increases). This indicates that when the cost of balancing feeding efficiency across both hosts is small, the evolutionary outcome favors reduced allocation toward feeding on host $h2$. In other words, weakening the trade-off drives the optimal feeding strategy toward less specialization on $h2$, thereby lowering the evolutionary singular strategy, as shown in Figure \ref{fig8}.

\begin{figure}[ht!]
	\centering
	\includegraphics[width=0.35\linewidth]{ 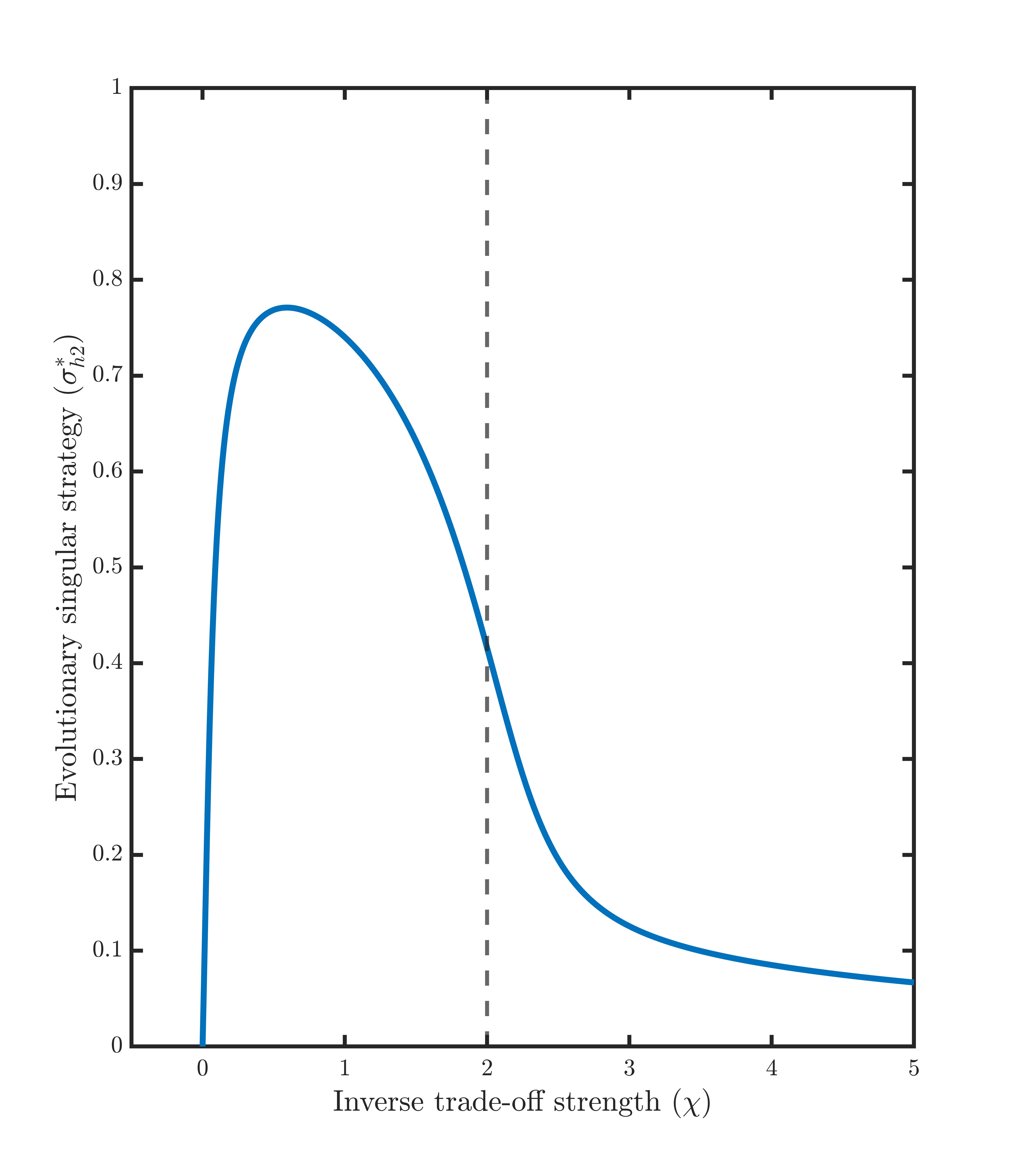}
	\caption{\textbf{ Dependence of evolutionary singular strategy of vector population on the trade-off strength.} The parameter values used were such that host encounter rates were equal and the conditions \eqref{condition} were satisfied. That is, $\mathcal{E}_{h1}=\mathcal{E}_{h2}$=1, $q_2=$0.1, $\tau_2$=1 day, $\tau_f$=0.2 day, $f_{rc}=4$ and rest parameters were as in Figure \ref{fig5:PIP}. Value of $\sigma_{h2}^*$ monotonically decreases as $\chi$ exceeds the value of 2 (dotted line).}
	\label{fig8}
\end{figure}

\section{Coupled Eco-Evolutionary Dynamics of Vector-Borne Disease}\label{sec-ecoepi}

In the preceding sections, we have characterized the epidemiological dynamics of VBD across host populations and explored the adaptive evolution of vector biting behavior in response to host defenses. However, the evolution of vector traits can directly influence disease transmission, while the prevalence of infection can, in turn, shape selection on vector behavior; the two processes are coupled. In this section, we investigate the feedback loop whereby vector evolution affects epidemiological outcomes, and these outcomes then influence evolutionary trajectories of vector biting behavior.

To capture the coupled dynamics of vector evolution and disease spread, we incorporate the adaptive biting trait $h2$ into the model system \eqref{req}. Figure \ref{fig7} shows how the evolution of the trait to bite $h2$ in the vector population can influence the long-term dynamics of VBD. First, the values of $\sigma_{h2}^*$ are calculated for trade-off strength ranging from high to low (Figure \ref{fig7}(a)). Then vector preference is found using the expression\[\alpha_v=\frac{\sigma_{h1} \mathcal{E}_{h1} N_{h2}}{\sigma_{h2}^* \mathcal{E}_{h2} N_{h1}} .\] 
For the derivation of this expression, see \ref{deri_vec_pref}. The corresponding values of $c, b_3, d_3$ are substituted in $K=\frac{b_3-d_3}{b_{31}}$ and in the model equations \eqref{req} to obtain the corresponding values of $R_0$ (Figure \ref{fig7}(b)), and equilibrium prevalence (Figure \eqref{fig7}(c)-(e)). It is evident that the value of $R_0$, the prevalence in $h1$, and the prevalence in the vector population rise as the trade-off strength decreases. This demonstrates that VBD among $h1$ can be reduced if the vector encounters both hosts at similar rates and the trade-off strength between specialization and succeeding against host defense is strong. Hence, interventions protecting the preferred host $h1$ can alter vector feeding patterns by increasing vector interaction with $h2$. This can result in a decrease in the prevalence of VBD in $h1$, but the combined prevalence in $h1$ and $h2,$ i.e., $y_1^*+y_2^*$ can increase, as illustrated in Figure \ref{fig7}(f). Overall, this framework provides a setting to examine how interventions targeting particular hosts can indirectly alter vector feeding behavior, thereby shifting disease transmission across hosts and influencing long-term disease dynamics in multi-host populations.\\

\begin{figure}[ht!]
\centering
\includegraphics[width=0.7\linewidth]{ 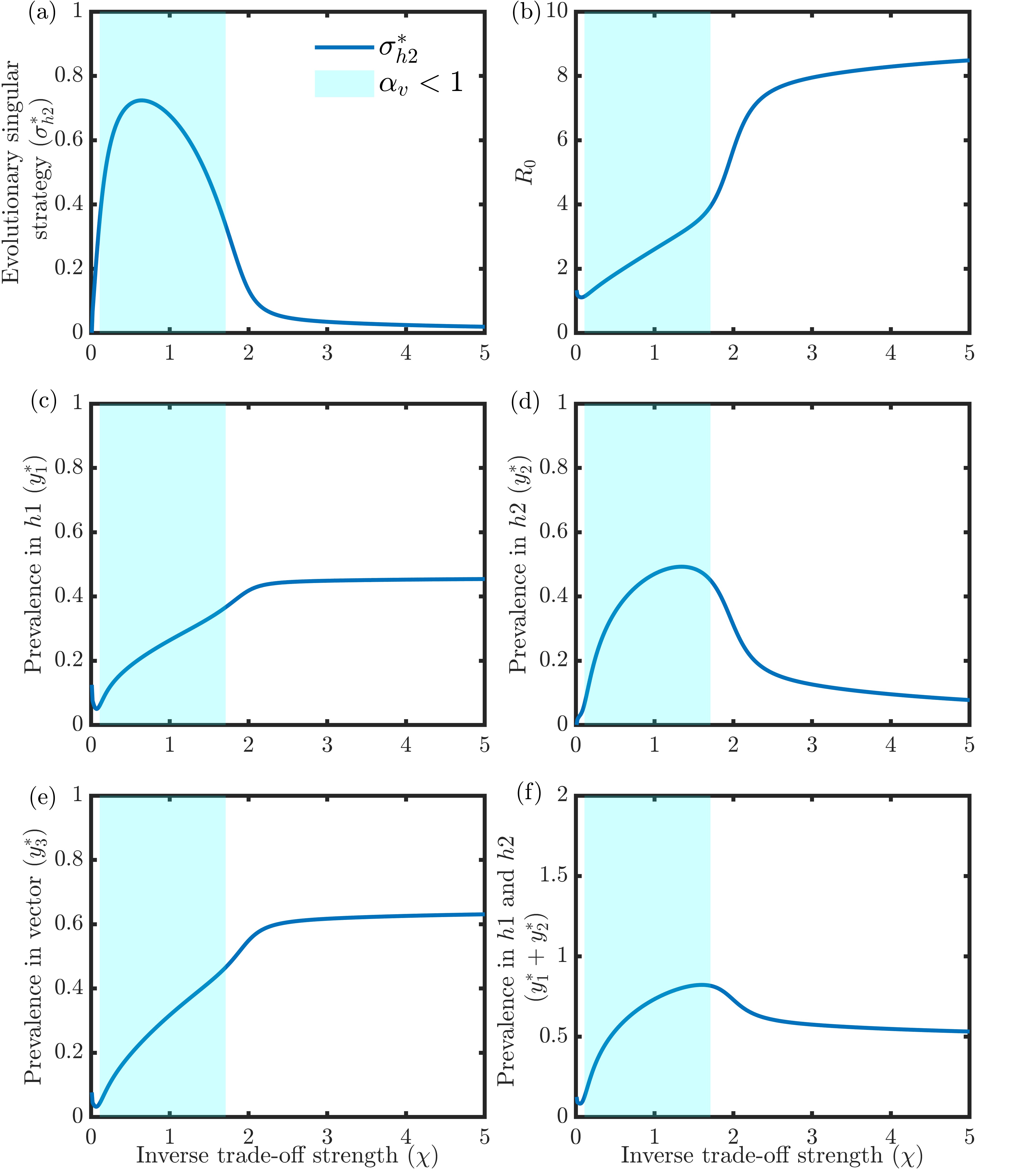}
\caption{ \textbf{The value of evolutionary singular point, $\mathbf{R_{0}}$, equilibrium prevalence in $\mathbf{h1}$, $\mathbf{h2}$, vector population, equilibrium prevalence in host population ($\mathbf{h1}$ and $\mathbf{h2}$) obtained for different $\mathbf{\chi}$ values.} The shaded region denotes the range of $\chi$ for which value of $\alpha_v<1$. For very weak values of trade-off strength, the prevalence decreases in $h2$ and increases in $h1$. Overall prevalence among the host population peaks in the shaded region (f). Parameter values used were $\mathcal{E}_{h1}$=1.2, $\mathcal{E}_{h2}$=1,  $\tau_2$=0.7 day and rest as in Figure \ref{fig5:PIP}.}
\label{fig7}
\end{figure}

We also investigate how host competence interacts with vector preference to shape disease outcome. Since host competence and the adaptation of vector biting behavior both influence the epidemiological outcomes, we simultaneously vary the competence of each host species and the vector preference, and compute the relative prevalence in the host $h1$ population for each combination. This allows us to identify under which conditions vector evolution {\em amplifies or mitigates} infection in the primary hosts, highlighting the intricate feedback loop between host traits, vector behavior, and disease dynamics. Figure \ref{Lfig9} illustrates the relationship between host competence $(\gamma=\frac{1}{\alpha_{vc}})$, vector preference $(\alpha_v)$ and disease prevalence of $h1$  ($y_1^*$). This highlights the importance of including vector preference and host competence while designing the public health policies for disease control. 

\begin{figure}[ht!]
\centering
\includegraphics[width=0.6\linewidth]{ 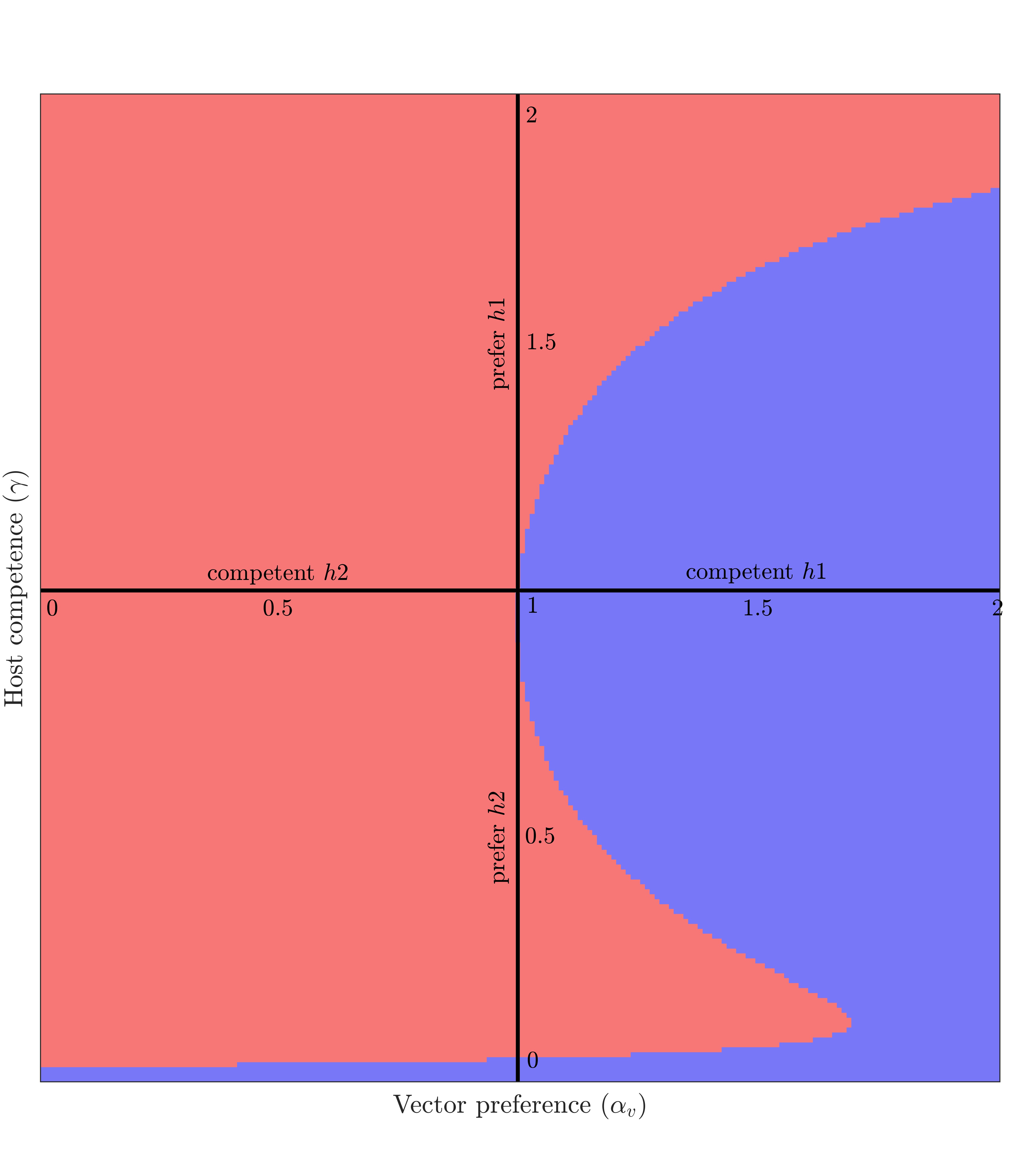}
\caption{\textbf{Behavior of $\mathbf{y_{1}^*-y_{1}^*(\gamma=1,\alpha_v=1)}$ as a function of $\mathbf{\alpha_v}$ and $\mathbf{\gamma}$.} Values are centered by subtracting $y_1^*$ obtained under conditions of no vector preference and equal transmission ability of the two hosts. Positive and negative deviations are coloured in red and blue to identify the amplification and dilution effects (increase or decrease relative to the reference value).  Amplification effect occurs when $h2$ is competent. Here $\mu_2$ is varied to vary $\gamma=\frac{\mu_2 + d_2}{\mu_1 +d_1}$. As $\dv{y_1^*}{\gamma}=\dv{ y_1^*}{\mu_2} \; \dv{ \mu_2}{ \gamma}<0$, $y_1^*$ decreases as $\gamma$ increases. Parameter $d_2=0.01$ and the rest were as in Table \ref{table1}.  }
\label{Lfig9}
\end{figure}

\section{Discussion}  \label{sec-con}
Understanding complex dynamics of multi host VBD is crucial from policy making viewpoint as interventions designed with the focus only on a single host species are known to shift vector preference and impact the broader disease ecology~\cite{gandy2022no,keesing2006effects}. We found that the effect of host population size on $R_0$ is not straightforward, which complements the findings by~\cite{Miller}. We demonstrated that an increase in host population size reduces the spread of VBD when the recovery rate for that host exceeds a certain threshold, which explains the previous results showing that an abundance of the low competent host population can reduce VBD transmission~\cite{keesing2006effects}. 

A key result of our analysis is the identification of a threshold, $R_{0c}$ - the minimum value of $R_0$ attained at the value of vector preference equal to the ratio of the rate at which $h2$ leaves the infectious class relative to $h1$. Recently, it has been shown that heterogeneous transmissibility can lead to major epidemics even at subcritical $R_0$~\cite{tuschhoff2025heterogeneity}. Our results complement this perspective by demonstrating that heterogeneity in host competence, together with vector feeding preferences, can also modulate epidemic outcomes. The threshold $R_{0c}$ characterizes this impact of vector preference on the equilibrium prevalence of $h1$ and marks a transition in the qualitative behavior of the system. We found that when \( h1 \) is a highly competent host, the prevalence in \( h1 \) increases steadily as vector preference \( \alpha_v \) shifts toward it. Interestingly, in this setting, the presence of a second host, \( h2 \), with lower competence, can help reduce the infection level in \( h1 \), particularly if some proportion of vector bites are redirected toward \( h2 \). The effect is more pronounced when \( h2 \) contributes little to vector infection, acting more as a biological buffer than a source. Such a concept of leveraging alternative host species to reduce human-vector interactions, known as zooprophylaxis, has significant empirical support as a promising approach in VBD management \cite{kemibala2020zooprophylaxis}. 

Our findings indicate that when $R_{0c}$ is near 1 and $\alpha_v$ is decreasing below $\alpha_{vc}$ then $y_1^*$ increases and attains a maximum at some point in $(0,\alpha_{vc})$. This suggests that protecting humans ($h1$) can concentrate bites on surrounding unprotected hosts and result in increased infections when $R_{0c}$ is low. This is because in such a case, the abundance of $h2$ can cause amplification of infected cases \cite{levine2017avian}. Moreover, our results also indicate that the introduction of secondary hosts like livestock in the presence of vectors that show moderate preference towards $h2$ can increase the transmission of VBD. These findings suggest that interventions designed to redirect vector feeding should be evaluated in the context of host competence, recovery rates, and vector behavior, as their epidemiological consequences may differ across systems. Our model simulations showed a decline in prevalence among $h1$, while prevalence in $h2$ and the vector increased. Hence, our approach also offers a framework to evaluate the effectiveness of implementing interventions such as ITN and vector repellents for abating prevalence among $h1$ in the presence of $h2$.

Studies incorporating vector preference into their transmission terms \cite{rivera2020relation,zahid2020decoys,marini2017exploring,simpson} have largely focused on its epidemiological consequences while overlooking its evolutionary dynamics.
Our results show that vector populations tend to evolve toward one of two extremes: a {\em specialist strategy} where vectors feed exclusively on \( h1 \), or a {\em generalist strategy} in which vectors consistently bite \( h2 \) whenever possible. The direction of evolution depends on the initial trait distribution, which separates the basins of attraction for these two strategies on the Pairwise Invasibility Plot (PIP). By examining how the singular strategy varies with trade-off strength, we found that the probability of a host being targeted by vectors depends not only on its abundance but also on the `cost' of specialization and the host's defensive capabilities. In particular, we find that a high feeding preference for \( h1 \) can be redirected toward \( h2 \) when the encounter rate with \( h2 \) equals or exceeds that of \( h1 \). We found that the likelihood of vectors attacking \( h2 \) decreases when the trade-off between host specialization and feeding success weakens. Interventions such as insecticide-treated nets or repellents can reduce encounters between vectors and \( h1 \), thereby strengthening the trade-off and effectively increasing the relative encounter rate with \( h2 \). This shift lowers the evolutionary preference \( \alpha_v \) for \( h1 \), and ultimately helps reduce infection prevalence in that host.

 For future work one can incorporate real-world data on host and vector populations, vector feeding patterns, and intervention outcomes for a more robust validation and refinement of our model predictions. Our framework highlights the potential for adaptive vector behavior and host heterogeneity to influence disease dynamics, suggesting that interventions targeting a single host or ignoring vector adaptation may have unintended consequences. With further development and integration of evolutionary and adaptive analyses, our modeling approach can help identify optimal strategies for vector control, guide public health decision-making, and improve the overall understanding of VBD dynamics across diverse ecological contexts.

\section*{Acknowledgements} \label{sec-ack}

\noindent This work is supported by the Science and Engineering Research Board (MTR/2020/000518 to AS).

\section*{Data Availability} \label{sec-data}

\noindent Data sharing is not applicable to this article as no new data were created or analyzed in this study.


\appendix

\section{Computation of $R_0$ using the next generation matrix method} \label{AppA}
\noindent The Next Generation Matrix (NGM) method is employed to determine the basic reproduction number, $R_0$, which is the expected number of secondary infections caused by a single infected individual in a fully susceptible population. To apply this method, we first decompose the Jacobian matrix evaluated at the disease-free solution ($y_{1}^*=0,z_{1}^*=0,y_{2}^*=0,y_{3}^*=0, M^*=K$) into two components: $T$ and $\Sigma$. The matrix $T$ represents the transmission terms, capturing the production of new infections, while $\Sigma$ accounts for the transitions within compartments, such as recovery, mortality, or immunity acquisition. This decomposition facilitates the derivation of the NGM and the subsequent calculation of $R_0$. 

Let $ c_1=\dfrac{\beta_{hv}c }{(\alpha_{v} 
N_{h1}+N_{h2})}$ and $c_2=\dfrac{\beta_{vh}c}{(\alpha_{v} N_{h1}+N_{h2})}$, then for system \eqref{req} we get the following,
\begin{align*}
T=   \begin{bmatrix}
	0&0&0&\alpha_v c_1 N_{h1}&0\\
	0&0&0&0&0\\
	0&0&0& c_1 N_{h2}&0\\
	c_2 \alpha_v K&0&c_2 K&0&0\\
    0&0&0&0&0&
\end{bmatrix}
\end{align*}
\begin{align*}
\Sigma=\begin{bmatrix}
	-(\mu_1+d_1)&0&0&0&0\\
	\mu_1&-(d_1+\delta_1)&0&0&0\\
	0&0&-(\mu_2+d_2)&0&0\\
	0&0&0&-d_3&0\\
    0&0&0&0&b_3-d_3-2b_{31}K
\end{bmatrix},
\end{align*}
Now we find the next generation matrix $\mathcal{K}$ by multiplying the transmission matrix, $T$, with the inverse of the transition matrix, $\Sigma^{-1}.$  
\begin{align*}
\mathcal{K}&=  -T\Sigma^{-1}=  \begin{bmatrix}
	0&0&0&-\alpha_v c_1 N_{h1}&0\\
	0&0&0&0&0\\
	0&0&0& -c_1 N_{h2}&0\\
	-c_2 \alpha_v K&0&-c_2 K&0&0\\
     0&0&0&0&0&
\end{bmatrix}
\begin{bmatrix} 
	\frac{-1}{(\mu_1+d_1)}&0&0&0&0\\
	\frac{-\mu_1}{(\mu_1+d_1)(\delta_1+d_1)}&\frac{-1}{(d_1+\delta_1)}&0&0&0\\
	0&0&\frac{-1}{(\mu_2+d_2)}&0&0\\
	0&0&0&\frac{-1}{d_3}&0\\
    0&0&0&0&\frac{1}{b_3-d_3-2b_{31} K}
\end{bmatrix},\\&
\mathcal{K}=\begin{bmatrix}
	0&0&0&\frac{\alpha_v c_1 N_{h1}}{d_3}&0\\
	0&0&0&0&0\\
	0&0&0&\frac{c_1 N_{h2}}{d_3}&0\\
	\frac{ c_2\alpha_v K}{(\mu_1+d_1)}&0&\frac{c_2 K}{(\mu_2+d_2)}&0&0\\
    0&0&0&0&0
\end{bmatrix}.
\end{align*}
The resulting matrix $\mathcal{K}$ indicates the expected number of secondary infections induced by an infected individual within each respective compartment during their infectious period. The basic reproduction number is then computed as the dominant eigenvalue (spectral radius) of this matrix. 

The spectral radius is
$\rho(\mathcal{K})=\sqrt{\frac{\beta_{hv}\beta_{vh}c^2 K}{(\alpha_v N_{h1}+N_{h2})^2d_3}\left(\frac{\alpha_{v}^{2} N_{h1}}{(d_1 +\mu_{1})} +\frac{N_{h2}}{(d_{2}+\mu_2)}\right)}.$
By verifying the condition for the existence of EE, we can write \eqref{Ro}.

\section{Proof of Theorem 1} \label{proof local stable DFE}
\begin{proof}
The Jacobian of the system \eqref{req} evaluated at the DFE $E_0=(0,0,0,0,M^*)$, is given by
\footnotesize
\setlength{\arraycolsep}{3.5pt} 
\medmuskip = 3mu 
\[ \hspace{-3cm}
{  J(E_0) }= 
\left( \begin{array}{ccccc} 
	-(\mu_1+d_1)&       0       &   0        & \frac{\alpha_{v}c \beta_{hv}N_{h1}}{(\alpha_{v} N_{h1}+N_{h2})}   &   0  \\
	\mu_{1}     &-(\delta_{1}+d_{1})       &      0                    &0 &   0   \\
	0      &0        &       -(\mu_2+d_{2})   &\frac{c\beta_{hv}N_{h2}}{(\alpha_{v} N_{h1}+N_{h2})} &   0  \\
	\frac{\beta_{vh}c\alpha_{v} M^*}{(\alpha_{v} N_{h1}+N_{h2})}&0 &      \frac{\beta_{vh}c M^*}{(\alpha_{v} N_{h1}+N_{h2})}           &-d_3 &   0  \\
     0       & 0       & 0       & 0       & b_3-d_3-2b_{31} M^*
\end{array} \right).
\]
\\
\normalsize
At $M^*=0$, the eigenvalues of $J(E_{00})$ are $\lambda_1=-(\delta_1+d_1)<0$, $\lambda_2=b_3-d_3$,  $\lambda_3=-(\mu_1+d_1)<0$, $\lambda_4=-(\mu_2+d_2)<0$, $\lambda_5=-d_3<0$. Hence, it is unstable since $b_3>d_3$, that is, the vector growth rate is positive.
At $M^*=K$, two of the eigenvalues of $J(E_{0K})$ are $\lambda_1=-(\delta_1+d_1)<0$, and $\lambda_2=-(b_3-d_3)<0$. The remaining three eigenvalues can be obtained from the equation 
\begin{equation} \label{3}
	\lambda^3+A_1\lambda^2+A_2 \lambda+A_3=0,
\end{equation}
where $A_1,A_2$ and $A_3$ are given by 
\begin{align*} 
	\begin{split}
		A_1&=\mu_2+d_2+\mu_1+d_1+d_3,\\
		A_2&=d_3(\mu_1+d_1)+(\mu_1+d_1+d_3)(\mu_2+d_2)-\frac{\beta_{hv}\beta_{vh}c^2 K (\alpha_v^2 N_{h1}+N_{h2})}{(\alpha_v N_{h1}+N_{h2})^2},\\
		A_3&=(1-R_0)d_3(\mu_1+d_1)(\mu_2+d_2).
	\end{split}
\end{align*}
Using the Routh Hurwitz criterion \cite{Stephen}, we find that the roots of the equation \eqref{3} will have a negative real part if $R_0<1$. 
Hence, the disease-free equilibrium is locally asymptotically stable if $R_0<1$. 
\end{proof}

\section{Proof of Theorem 2} \label{proof local stable}
\begin{proof}
Since the vector population equation is decoupled and its positive equilibrium $M=K$ is asymptotically stable, the stability of the endemic equilibrium is determined by the reduced four-dimensional subsystem obtained by setting $M=K.$ Let $Y_1=y_1-y_1^{*}$, $Z_1=z_1-z_1^{*}$, $Y_2=y_2 - y_2^{*}$, and $Y_3=y_3-y_3^{*}$ are small perturbations in $y_1,y_2,y_3$ and $z_1$ around $E^{*}$. Consider, 
\begin{align*}
	&f_{11}=-(\alpha_{v} c_1 y_3^{*} +\mu_{1} +d_{1}),\;
	f_{13}=\alpha_{v} c_1 (1-y_1^{*}-z_1^{*}),\;
	f_{14}=-\alpha_{v} c_1 y_3^{*},\\
	&f_{22}=-(c_1 y_3^{*} +\mu_2+d_2),\;
	f_{23}=c_1 (1-y_2^{*}),\;
	f_{31}=c_2 \alpha_{v} N_{h1}(1-y_3^{*}),\\
	&f_{32}=c_2 N_{h2} (1-y_3^{*}),\;
	f_{33}=-c_2 (\alpha_{v} N_{h1}\; y_1^{*}+N_{h2}\; y_2^{*} )-d_3,\;
	f_{41}=\mu_{1},\;
	f_{44}=-(\delta_{1}+d_1).
\end{align*}
Then the linearised system corresponding to EE,  $E^{*}$ is:
$$
\begin{bmatrix}
	\dot{Y_1}\\
	\dot{Y_2}\\
	\dot{Y_3}\\
	\dot{Z_1}\\
\end{bmatrix}
=
\left[ \begin{array}{cccc} 
	f_{11}&0&f_{13}&f_{14} \\
	0&f_{22}&f_{23}&0\\
	f_{31}&f_{32}&f_{33}&0\\
	f_{41}&0&0&f_{44}
\end{array} \right]
\begin{bmatrix}
	Y_1\\
	Y_2\\
	Y_3\\
	Z_1\\
\end{bmatrix}.
$$
Now, consider the positive definite function $$W=\frac{1}{2} Y_1^2+\frac{w_1}{2} Z_1^2+\frac{w_2}{2} Y_2^2+\frac{w_3}{2} Y_3^2,$$ 
where $w_i\;(i=1,2,3) > 0$ are chosen so that $\dot{W}$ is negative definite, 
\begin{equation*}
	\begin{aligned} 
		\dot{W}&=\alpha_v c_1 (1-y_1^{*}-z_1^{*})Y_3Y_1+w_3c_2\alpha_v N_{h1}(1-y_3^{*})Y_1Y_3+w_2c_1(1-y_2^{*})Y_3Y_2\\&+w_3 c_2 N_{h2}(1-y_3^{*})Y_2Y_3-\alpha_vc_1y_3^{*}Z_1Y_1+w_1\mu_1Y_1Z_1-(\alpha_vc_1y_3^{*}+\mu_1+d_1)Y_1^2\\&-w_2(c_1y_3^{*}+\mu_2+d_2)Y_2^2-w_3(c_2(\alpha_vN_{h1}y_1^{*}+N_{h2}y_2^{*})+d_3)Y_3^2-w_1(d_1+\delta_1)Z_1^2.
	\end{aligned}
\end{equation*}
Set $w_1=\frac{\alpha_v c_1 y_3^{*}}{\mu_1}$, then we can derive that $\dot{W}$ is negative definite if the following are true:
\begin{align}
	\frac{2\alpha_v^2c_1^2(1-y_1^{*}-z_1^{*})^2}{d_3(\mu_1+d_1)}&<w_3<\frac{d_3(\mu_1+d_1)}{2c_2^2\alpha_v^2N_{h1}^2(1-y_3^{*})^2},\label{local1}\\
	\frac{w_3 2 c_2^2 N_{h2}^2 (1-y_3^{*})^2}{(\mu_2+d_2)d_3}&<w_2<\frac{(\mu_2+d_2)d_3 w_3}{2c_1^2(1-y_2^{*})^2}\label{local2}.
\end{align}
From here, it follows that a positive $w_3$ can be chosen if the following inequality is satisfied:
\begin{equation*}
	2 c_1 c_2 \alpha_v^2 N_{h1}<d_3 (\mu_1+d_1).
\end{equation*}
Again, by substituting \eqref{local1} in \eqref{local2}, a positive $w_2$ can be chosen if the following inequality is satisfied:
\begin{equation*}
	2^2 \alpha_v^2 c_1^2 c_2^2 N_{h1} N_{h2}<d_3^2 (\mu_1+d_1)(\mu_2+d_2).
\end{equation*}
By substituting $c_1$ and $c_2$ in the above two inequalities, we obtain the desired result \eqref{5}.
\end{proof}

\section{Proof of Theorem 3}\label{proof global stable}
\begin{proof}
Since the vector population equation is decoupled and its its value at $E^*$ is given by $M=K$, we analyze the reduced four-dimensional system obtained by setting $M=K$. To establish the global asymptotic stability of the $E^*$, consider the Lyapunov function 
\begin{equation*}
	\begin{aligned}
		U&= \frac{1}{2}(y_1-y_1^\ast)^2 + \frac{u_1}{2}(z_1-z_1^\ast)^2+ \frac{u_2}{2}(y_2-y_2^\ast)^2 + \frac{u_3}{2}(y_3-y_3^\ast)^2,
	\end{aligned}
\end{equation*}
where the constants $u_1,u_2$, and $u_3$ are chosen so that the derivative of $U$ along solutions of the reduced system is negative definite.
\begin{equation*}
	\begin{aligned}
		\dot{U}&=\alpha_v c_1(1-y_1^{*}-z_{1})(y_1-y_1^{*})(y_3-y_3^{*})+u_3c_2\alpha_vN_{h1}(1-y_3^*)(y_1-y_1^{*})(y_3-y_3^{*})\\&+u_2c_1(1-y_2^*)(y_3-y_3^{*})(y_2-y_2^{*})+u_3c_2N_{h2}(1-y_3^*)(y_3-y_3^{*})(y_2-y_2^{*})-(\mu_1+d_1)(y_1-y_1^{*})^2\\&-u_2 (d_2 + \mu_2)(y_2-y_2^{*})^2-u_3d_3(y_3-y_3^{*})^2-\alpha_v c_1 y_3(y_1-y_1^{*})^2-u_1(d_1+\delta_1)(z_1-z_1^{*})^2\\&-u_2c_1 y_3(y_2-y_2^{*})^2-u_3c_2\alpha_vN_{h1}y_1(y_3-y_3^{*})^2-u_3c_2N_{h2}y_2(y_3-y_3^*)^2.
	\end{aligned}
\end{equation*}
By selecting $u_1$ as $u_1= \frac{\alpha_v c_1 y_{3}^{*}}{\mu_1}$, $\dot{U}$ becomes negative definite if the following conditions hold true:

\begin{align}   
	\frac{\alpha_v^2 c_1^2 (1-y_1^{*}-z_{1})^2}{(\mu_1+d_1)d_3}& < u_3 <  \frac{(\mu_1+d_1)d_3}{2 \alpha_v^2 c_2^2 N_{h1}^2 (1-y_3^*)^2},\label{global1}\\
	\frac{u_3 2 c_2^2 {N_{h2}}^2 (1-y_3^*)^2 }{d_3(\mu_2+d_2)}& < u_2 < \frac{u_3 d_3 (d_2 + \mu_2)}{c_1^2(1-y_2^*)^2}\label{global2}.
\end{align}
\noindent Since $z_1<1$, we can choose a positive $u_3$ if the following hold
\begin{align*}
	2\alpha_v^2 c_1 c_2 N_{h1}<(\mu_1+d_1)d_3.
\end{align*}
Now using \eqref{global1} in \eqref{global2}  we can choose positive $u_2$ if the following hold
\begin{equation*}
2^2 \alpha_v^2 c_1^2 c_2^2 N_{h1} N_{h2} <d_3^2 (\mu_2+d_2)(\mu_1+d_1).
\end{equation*}
Hence, we get sufficient condition \eqref{6} for global stability in $\Gamma$.
\end{proof}

\section{Proof of Proposition 1}\label{pf:prop_5.1}
\begin{proof}
Differentiating $\psi$ with respect to $\sigma_{h2(m)}$, we obtain
\begin{align*}
\textstyle \frac{\partial \psi }{\partial \sigma_{h2(m)}}&\textstyle=\frac{1}{\left( \tau_f+\tau_2\left(1-p_2 e^{-\mathcal{E}(1+\sigma_{h2(m)})}\right)\right) } \left[ \frac{(1-p_2)\mathcal{E}e^{-\mathcal{E}(1+\sigma_{h2(m)})}}{\left(1-e^{-\mathcal{E}(1+\sigma_{h2(m)})}\right)}+
\frac{\nu \left(\sigma_{h2(m)}^{\chi -1} (\sigma_{h2(m)}-\chi)+\chi \sigma_{h2(m)}-1\right) \left(1-p_2 e^{-\mathcal{E}(1+\sigma_{h2(m)})}\right)}{(1+\sigma_{h2(m)}) \left[(1+\sigma_{h2(m)})^{\chi+1}-\nu \sigma_{h2(m)}\left(1+\sigma_{h2(m)}^{\chi-1}\right)\right]}  \right] \\&\textstyle +\left(f_{rc}+\log(q_1 q_2) \right) \frac{\tau_f p_2 \mathcal{E}e^{-\mathcal{E}(1+\sigma_{h2(m)})}}{\left( \tau_f+\tau_2\left(1-p_2 e^{-\mathcal{E}(1+\sigma_{h2(m)})}\right)\right)^2}.
\end{align*}
Now, at singular strategy, i.e., at $\sigma_{h2}^*$, the fitness gradient $\left[\frac{\partial\psi_{\sigma_{h2(r)}}(\sigma_{h2(m)})}{\partial\sigma_{h2(m)}}\right]_{\sigma_{h2(r)}=\sigma_{h2(m)}^*,\atop \sigma_{h2(m)}=\sigma_{h2(m)}^*}=0$, which implies
\begin{equation}
\begin{aligned} \label{chi_derivative}
	&\left( \tau_f+\tau_2\left(1-p_2 e^{-\mathcal{E}(1+\sigma_{h2}^*)}\right)\right) \left[ \frac{(1-p_2)\mathcal{E}e^{-\mathcal{E}(1+\sigma_{h2}^*)}}{\left(1-e^{-\mathcal{E}(1+\sigma_{h2}^*)}\right)}+\frac{\nu \left(\sigma_{h2}^{\chi-1} (\sigma_{h2}^*-\chi)+\chi\sigma_{h2}^*-1\right)\left(1-p_2 e^{-\mathcal{E}(1+\sigma_{h2}^*)}\right)}{\left(1+\sigma_{h2}^*\right)\left((1+\sigma_{h2}^{*})^{\chi+1}-\nu(\sigma_{h2}^{*}+(\sigma_{h2}^{*})^{\chi})\right)}\right] \\
	& + \left(f_{rc}+\log(q_1 q_2) \right) \tau_f p_2 \mathcal{E}e^{-\mathcal{E}(1+\sigma_{h2}^*)}=0.
\end{aligned}
\end{equation}

Moreover, $\frac{\partial}{\partial\chi}$ $\left[\frac{\partial\psi_{\sigma_{h2(r)}}(\sigma_{h2(m)})}{\partial\sigma_{h2(m)}}\right]_{\sigma_{h2(r)}=\sigma_{h2(m)}^*,\atop \sigma_{h2(m)}=\sigma_{h2(m)}^*}=0$ and therefore using \eqref{chi_derivative}, one can obtain the following

\begin{align*}
&\textstyle \frac{\partial \sigma_{h2}^*}{\partial \chi} \frac{\nu \chi	\left( \tau_f+\tau_2\left(1-p_2 e^{-\mathcal{E}(1+\sigma_{h2}^{*})}\right)\right) \left(1-p_2 e^{-\mathcal{E}(1+\sigma_{h2}^*)}\right) }{(1+\sigma_{h2}^*)\left((1+\sigma_{h2}^{*})^{\chi+1}-\nu(\sigma_{h2}^*+(\sigma_{h2}^{*})^{\chi}) \right) } \left((\sigma_{h2}^{*})^{\chi-1}+1-(\chi-1)(\sigma_{h2}^{*})^{\chi-2}\right)
+	\\
&\textstyle \frac{\partial \sigma_{h2}^*}{\partial \chi} \frac{\nu ((\sigma_{h2}^{*})^{\chi} -1+\chi (\sigma_{h2}^{*}-(\sigma_{h2}^{*})^{\chi-1}) \left( \tau_f+\tau_2\left(1-p_2 e^{-\mathcal{E}(1+\sigma_{h2}^*)}\right)\right) }{(1+\sigma_{h2}^{*})\left((1+\sigma_{h2}^{*})^{\chi+1}-\nu(\sigma_{h2}^*+(\sigma_{h2}^{*})^{\chi}) \right) } \Bigg[2 p_2 \mathcal{E} e^{-\mathcal{E}(1+\sigma_{h2}^{*})} \\
&\textstyle -\left(1-p_2 e^{-\mathcal{E}(1+\sigma_{h2}^{*})}\right)\bigg( \frac{1}{(1+\sigma_{h2}^*)} +\frac{(\chi+1)(1+\sigma_{h2}^*)^\chi-\nu (1+\chi (\sigma_{h2}^{*})^{\chi-1})}{\left((1+\sigma_{h2}^{*})^{\chi+1}-\nu(\sigma_{h2}^*+(\sigma_{h2}^{*})^{\chi}) \right) }\bigg)  \Bigg] \\
&\textstyle + \frac{\partial \sigma_{h2}^*}{\partial \chi} \frac{(1-p_2)\mathcal{E}^2}{\left(1-e^{-\mathcal{E}(1+\sigma_{h2}^{*})}\right)}
\left( \tau_f+\tau_2\left(1-p_2 e^{-\mathcal{E}(1+\sigma_{h2}^{*})}\right)\right) \left[ \frac{p_2 e^{-2\mathcal{E}(1+\sigma_{h2}^{*})}}{\left(1-p_2  e^{-\mathcal{E}(1+\sigma_{h2}^{*})}\right)}-\frac{e^{-\mathcal{E}(1+\sigma_{h2}^{*})}}{\left(1- e^{-\mathcal{E}(1+\sigma_{h2}^{*})}\right)}\right]\\
&\textstyle	-\frac{\partial \sigma_{h2}^*}{\partial \chi}\left(f_{rc}+\log(q_1 q_2) \right) \tau_f p_2 \mathcal{E}^2e^{-\mathcal{E}(1+\sigma_{h2}^{*})} = \tau_f p_2 \mathcal{E} e^{-\mathcal{E}(1+\sigma_{h2}^{*})} \frac{\nu \left( \sigma_{h2}^{*} \log(1-\varphi) +\sigma_{h2}^{*\chi} \log(\varphi)\right)}{\left((1+\sigma_{h2}^{*})^{\chi+1}-\nu(\sigma_{h2}^{*}+\sigma_{h2}^{*\chi})\right)}\\
&\textstyle + \frac{\nu \left( \tau_f+\tau_2\left(1-p_2 e^{-\mathcal{E}(1+\sigma_{h2}^*)}\right)\right)  \left(1-p_2  e^{-\mathcal{E}(1+\sigma_{h2}^{*})}\right) }{(1+\sigma_{h2}^{*})\left((1+\sigma_{h2}^{*})^{\chi+1}-\nu(\sigma_{h2}^{*}+(\sigma_{h2}^{*})^{\chi})\right)} \Bigg[(\sigma_{h2}^*)^{\chi-1}-\sigma_{h2}^*+(\chi-\sigma_{h2}^*)(\sigma_{h2}^*)^{\chi-1}\log(\sigma_{h2}^*)+ \\
&\textstyle \frac{\left((\sigma_{h2}^{*})^{\chi -1} (\sigma_{h2}^{*}-\chi)+\chi \sigma_{h2}^{*}-1\right) \left( (1+\sigma_{h2}^{*})^{\chi+1} \log((1+\sigma_{h2}^{*}))-\nu \sigma_{h2}^{*} \log(\sigma_{h2}^{*})\right)}{\left((1+\sigma_{h2}^{*})^{\chi+1}-\nu(\sigma_{h2}^{*}+(\sigma_{h2}^{*})^{\chi})\right)}\Bigg].
\end{align*}

Since the term $(\sigma_{h2}^*)^{\chi-1}-\sigma_{h2}^*+(\chi-\sigma_{h2}^*)(\sigma_{h2}^*)^{\chi-1}\log(\sigma_{h2}^*)$  is always negative for $\chi>2$ and if we consider the condition (i) is true, i.e., $(\sigma_{h2}^{*})^{\chi -1} (\sigma_{h2}^{*}-\chi)+\chi \sigma_{h2}^{*}-1\le 0$, then the right hand side is negative. Therefore, it follows that $\frac{\partial \sigma_{h2}^*}{\partial \chi}$ must be negative if the followings are satisfied

\begin{equation}
\begin{aligned} \label{condi_1}
	&\textstyle (\sigma_{h2}^{*})^{\chi-1}+1-(\chi-1)(\sigma_{h2}^{*})^{\chi-2}\ge 0,\\
	& \Bigg[2 p_2 \mathcal{E} e^{-\mathcal{E}(1+\sigma_{h2}^{*})} -\left(1-p_2 e^{-\mathcal{E}(1+\sigma_{h2}^{*})}\right)\bigg( \frac{1}{(1+\sigma_{h2}^*)} +\frac{(\chi+1)(1+\sigma_{h2}^*)^\chi-\nu (1+\chi (\sigma_{h2}^{*})^{\chi-1})}{\left((1+\sigma_{h2}^{*})^{\chi+1}-\nu(\sigma_{h2}^*+(\sigma_{h2}^{*})^{\chi}) \right) }\bigg)  \Bigg]\le 0,\\
	& \textstyle -\left(f_{rc}+\log(q_1 q_2) \right) \tau_f p_2>\frac{\left( \tau_f+\tau_2\left(1-p_2 e^{-\mathcal{E}(1+\sigma_{h2})}\right)\right)}{\left(1- e^{-\mathcal{E}(1+\sigma_{h2}^{*})}\right)^2}(1-p_2).
\end{aligned}
\end{equation}

Finally, if the given conditions \eqref{condition} are true, then these imply the inequalities in \eqref{condi_1}. Hence $\frac{\partial \sigma_{h2}^*}{\partial \chi}<0$. This completes the proof.
\end{proof}


 \section{ Derivation of expression for vector preference } \label{deri_vec_pref}
Here we present how vector preference is obtained in the case of the evolution of the trait to bite $h2$. Let the parameters $K_1$ and $K_2$ represent the number of bites for $h1$ and $h2$, then the vector preference is given as 
\[\alpha_v=\frac{K_1/K_2}{N_{h1}/N_{h2}}=\frac{K_1}{K_2} \times \frac{N_{h2}}{N_{h1}}.\]
By multiplying the numerator and denominator with $(K_1+K_2)$, we obtain the following
\[	\alpha_v =\frac{K_1/(K_1+K_2)}{K_2/(K_1+K_2)} \times \frac{N_{h2}}{N_{h1}}.\]
This is equivalent to the following,\\
\begin{align*}
\alpha_v &	=\frac{\text{Probability that } h1 \text{ is bitten}}{\text{Probability that } h2 \text{ is bitten}} \times \frac{N_{h2}}{N_{h1}},\\
&=\frac{(1-\varphi)}{\varphi} \times \frac{N_{h2}}{N_{h1}},\\
&=\frac{\frac{\sigma_{h1}\mathcal{E}_{h1}}{\sigma_{h1} \mathcal{E}_{h1}+\sigma_{h2} \mathcal{E}_{h2}}}{\frac{\sigma_{h2}\mathcal{E}_{h2}}{\sigma_{h1} \mathcal{E}_{h1}+\sigma_{h2} \mathcal{E}_{h2}}} \times \frac{N_{h2}}{N_{h1}}.
\end{align*}
Consequently, the vector preference is given as follows
\[ \alpha_v=\frac{\sigma_{h1} \mathcal{E}_{h1} N_{h2}}{\sigma_{h2}^* \mathcal{E}_{h2} N_{h1}}. \]

\section{Numerical validation of the stability analysis for the endemic equilibrium $E^{*}$}
\noindent The basic reproduction number for parameter set in Figure \ref{fig1} is $R_0 = 3.2541>1$. The eigenvalue analysis reveals that all eigenvalues have negative real parts, confirming the local asymptotic stability of $E^{*}$. Moreover, the global stability conditions are also satisfied for these parameter values. The phase portrait illustrates that all the trajectories originating within the region of attraction converge to the endemic equilibrium, further supporting the global stability of $E^{*}.$

\begin{figure}[ht!]
\centering
\includegraphics[width=0.6\linewidth]{ 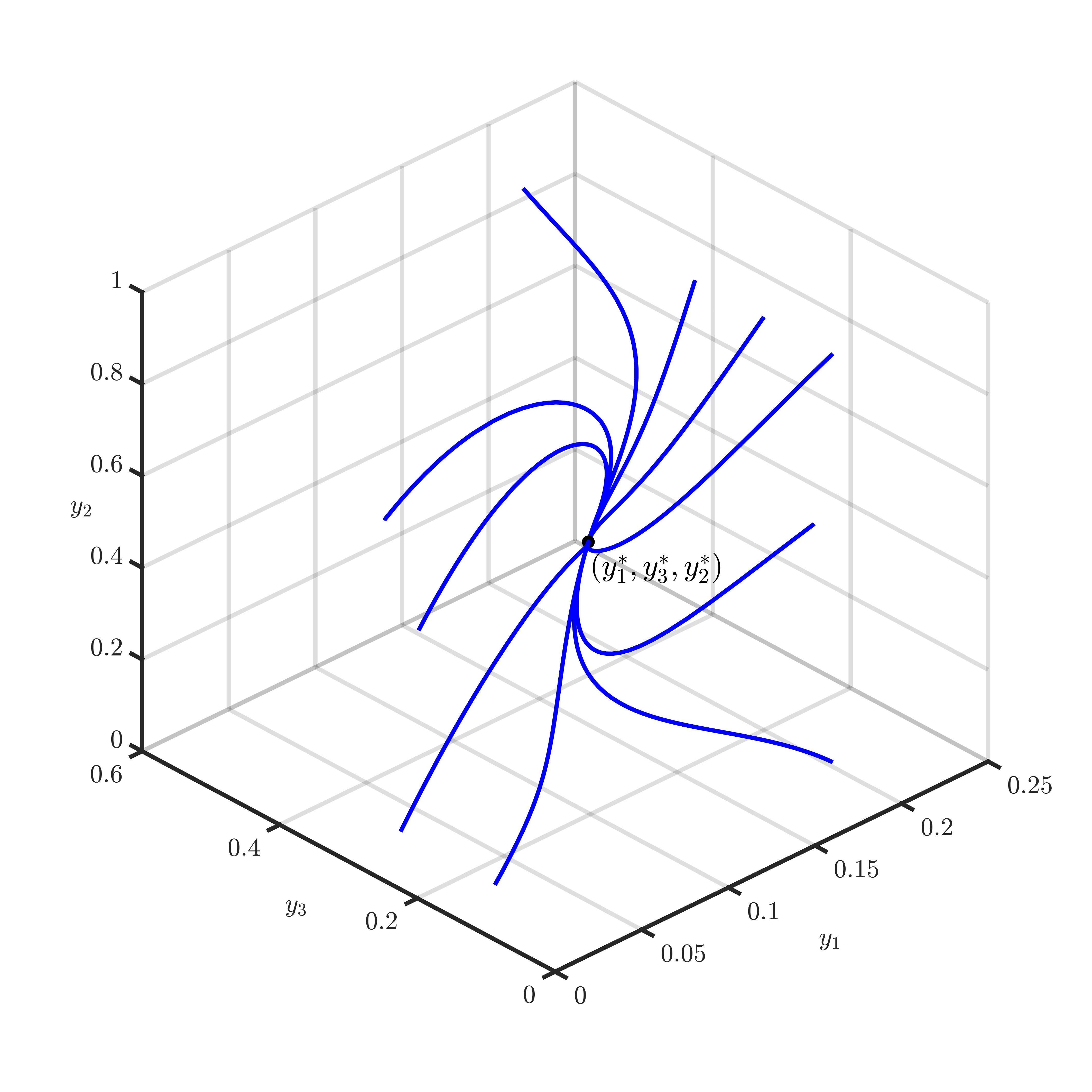}
\caption{\textbf{Nonlinear stability of $\mathbf{E^{*}}$ in $\mathbf{y_{1}-y_3-y_{2}}$ phase space}. Parameter $\alpha_v=0.1$ and rest were as in Table \ref{table1}.}
\label{fig1}
\end{figure}

 \bibliography{2hVBD_ref}

@article{Miller,
	author = {Ezer Miller and Amit Huppert},
	year = {2013},
	pages = {e80279},
	title = {The Effects of Host Diversity on Vector-Borne Disease: The Conditions under Which Diversity Will Amplify or Dilute the Disease Risk},
	volume = {8},
	journal = {PLoS One}}

@article{Macdonald,
	author = {Macdonald, G},
	year = {1956},
	title = {Epidemiological basis of malaria control},
	journal = {Bulletin of the World Health Organization},
	volume = {15},
	pages={613-626},
}

@article{simpson,
	title={Vector host-feeding preferences drive transmission of multi-host pathogens: West Nile virus as a model system},
	author={J. E. Simpson and P. J. Hurtado and J. Medlock and G. Molaei and T. G. Andreadis and A. P. Galvani and M. A. Diuk-Wasser},
	journal={Proceedings of the Royal Society B: Biological Sciences},
	volume={279},
	pages={925-933},
	year={2012},
}

@book{Stephen,
	title={Introduction to applied nonlinear dynamical systems and chaos},
	author={Stephen Wiggins},
	year={1990},
	publisher={Springerr-Verlag},
	address={New York}
}

@book{ross1911prevention,
	title={The prevention of malaria},
	author={Ross, Ronald},
	year={1911},
	publisher={John Murray},
	address={London}
}

@article{diekmann2010construction,
	title={The construction of next-generation matrices for compartmental epidemic models},
	author={Diekmann, Odo and Heesterbeek, JAP and Roberts, Michael G},
	journal={Journal of the Royal Society Interface},
	volume={7},
	pages={873-885},
	year={2010},
	publisher={The Royal Society},
}

@article{chitnisR_0,
	title={Determining important parameters in the spread of malaria through the sensitivity analysis of a mathematical model},
	author={Chitnis, Nakul and Hyman, James M and Cushing, Jim M},
	journal={Bulletin of Mathematical Biology},
	volume={70},
	pages={1272-1296},
	year={2008},
	publisher={Springer},
}

@article{martinez2021differential,
	title={Differential attraction in mosquito-human interactions and implications for disease control},
	author={Martinez, Julien and Showering, Alicia and Oke, Catherine and Jones, Robert T and Logan, James G},
	journal={Philosophical Transactions of the Royal Society B},
	volume={376},
	pages={20190811},
	year={2021},
	publisher={The Royal Society},
}

@article{keesing2006effects,
	title={Effects of species diversity on disease risk},
	author={Keesing, Felicia and Holt, Robert D and Ostfeld, Richard S},
	journal={Ecology Letters},
	volume={9},
	pages={485-498},
	year={2006},
	publisher={Wiley Online Library},
}

@article{chen2022host,
	title={Host competence, interspecific competition and vector preference interact to determine the vector-borne infection ecology},
	author={Chen, Lifan and Chen, Shiliang and Kong, Ping and Zhou, Liang},
	journal={Frontiers in Ecology and Evolution},
	volume={10},
	pages={993844},
	year={2022},
	publisher={Frontiers},
}

@article{rivera2020relation,
	title={The relation between host competence and vector-feeding preference in a multi-host model: Chagas and Cutaneous Leishmaniasis},
	author={Rivera, Rocio Caja and Bilal, Shakir and Michael, Edwin},
	journal={Mathematical Biosciences and Engineering},
	volume={17},
	pages={5561-5583},
	year={2020},
}

@article{gandy2022no,
	title={No net effect of host density on tick-borne disease hazard due to opposing roles of vector amplification and pathogen dilution},
	author={Gandy, Sara and Kilbride, Elizabeth and Biek, Roman and Millins, Caroline and Gilbert, Lucy},
	journal={Ecology and Evolution},
	volume={12},
	pages={e9253},
	year={2022},
	publisher={Wiley Online Library},
}

@article{rock2015age,
	title={Age-and bite-structured models for vector-borne diseases},
	author={Rock, Kat S and Wood, David A and Keeling, Matthew James},
	journal={Epidemics},
	volume={12},
	pages={20-29},
	year={2015},
	publisher={Elsevier}
}

@article{yakob2010modelling,
	title={Modelling knowlesi malaria transmission in humans: vector preference and host competence},
	author={Yakob, Laith and Bonsall, Michael B and Yan, Guiyun},
	journal={Malaria Journal},
	volume={9},
	pages={1-7},
	year={2010},
	publisher={BioMed Central}
}

@article{marini2017exploring,
	title={Exploring vector-borne infection ecology in multi-host communities: A case study of West Nile virus},
	author={Marini, Giovanni and Ros{\'a}, Roberto and Pugliese, Andrea and Heesterbeek, Hans},
	journal={Journal of Theoretical Biology},
	volume={415},
	pages={58-69},
	year={2017},
	publisher={Elsevier}
}

@article{chamchod2011analysis,
	title={Analysis of a vector-bias model on malaria transmission},
	author={Chamchod, Farida and Britton, Nicholas F},
	journal={Bulletin of Mathematical Biology},
	volume={73},
	pages={639-657},
	year={2011},
	publisher={Springer}
}

@article{kemibala2020zooprophylaxis,
	title={A zooprophylaxis strategy using L-lactic acid (Abate) to divert host-seeking malaria vectors from human host to treated non-host animals},
	author={Kemibala, Elison E and Mafra-Neto, Agenor and Dekker, Teun and Saroli, Jesse and Silva, Rodrigo and Philbert, Anitha and Nghabi, Kija and Mboera, Leonard EG},
	journal={Malaria Journal},
	volume={19},
	pages={1-7},
	year={2020},
	publisher={BioMed Central}
}

@article{levine2017avian,
	title={Avian species diversity and transmission of West Nile virus in Atlanta, Georgia},
	author={Levine, Rebecca S and Hedeen, David L and Hedeen, Meghan W and Hamer, Gabriel L and Mead, Daniel G and Kitron, Uriel D},
	journal={Parasites \& Vectors},
	volume={10},
	pages={1-12},
	year={2017},
	publisher={BioMed Central}
}

@article{ravigne2009live,
	title={Live where you thrive: joint evolution of habitat choice and local adaptation facilitates specialization and promotes diversity},
	author={Ravign{\'e}, Virginie and Dieckmann, Ulf and Olivieri, Isabelle},
	journal={The American Naturalist},
	volume={174},
	pages={E141-E169},
	year={2009},
	publisher={The University of Chicago Press}
}

@article{le2007elaborated,
	title={An elaborated feeding cycle model for reductions in vectorial capacity of night-biting mosquitoes by insecticide-treated nets},
	author={Le Menach, Arnaud and Takala, Shannon and McKenzie, F Ellis and Perisse, Andre and Harris, Anthony and Flahault, Antoine and Smith, David L},
	journal={Malaria Journal},
	volume={6},
	pages={1-12},
	year={2007},
	publisher={Springer}
}

@article{stone2018evolution,
	title={Evolution of host preference in anthropophilic mosquitoes},
	author={Stone, Chris and Gross, Kevin},
	journal={Malaria Journal},
	volume={17},
	pages={1-11},
	year={2018},
	publisher={Springer}
}

@article{geritz1997dynamics,
	title={Dynamics of adaptation and evolutionary branching},
	author={Geritz, Stefan AH and Metz, Johan AJ and Kisdi, {\'E}va and Mesz{\'e}na, G{\'e}za},
	journal={Physical Review Letters},
	volume={78},
	pages={2024},
	year={1997},
	publisher={APS}
}

@article{zahid2020decoys,
	title={Decoys and dilution: the impact of incompetent hosts on prevalence of Chagas disease},
	author={Zahid, Mondal Hasan and Kribs, Christopher M},
	journal={Bulletin of Mathematical Biology},
	volume={82},
	pages={41},
	year={2020},
	publisher={Springer}
}

@article{anwar2024mathematical,
	title={Mathematical models of Plasmodium vivax transmission: A scoping review},
	author={Anwar, Md Nurul and Smith, Lauren and Devine, Angela and Mehra, Somya and Walker, Camelia R and Ivory, Elizabeth and Conway, Eamon and Mueller, Ivo and McCaw, James M and Flegg, Jennifer A and others},
	journal={PLoS Computational Biology},
	volume={20},
	pages={e1011931},
	year={2024},
	publisher={Public Library of Science San Francisco, CA USA}
}

@article{bilal2020complexity,
	title={Complexity and critical thresholds in the dynamics of visceral leishmaniasis},
	author={Bilal, Shakir and Caja Rivera, Rocio and Mubayi, Anuj and Michael, Edwin},
	journal={Royal Society Open Science},
	volume={7},
	pages={200904},
	year={2020},
	publisher={The Royal Society}
}

@article{wu2023spatial,
	title={Spatial dynamics of malaria transmission},
	author={Wu, Sean L and Henry, John M and Citron, Daniel T and Mbabazi Ssebuliba, Doreen and Nakakawa Nsumba, Juliet and S{\'a}nchez C, H{\'e}ctor M and Brady, Oliver J and Guerra, Carlos A and Garc{\'\i}a, Guillermo A and Carter, Austin R and others},
	journal={PLoS Computational Biology},
	volume={19},
	pages={e1010684},
	year={2023},
	publisher={Public Library of Science San Francisco, CA USA}
}

@article{Althouse,
	author = {Althouse, Benjamin M. AND Lessler, Justin AND Sall, Amadou A. AND Diallo, Mawlouth AND Hanley, Kathryn A. AND Watts, Douglas M. AND Weaver, Scott C. AND Cummings, Derek A. T.},
	journal = {PLoS Neglected Tropical Diseases},
	publisher = {Public Library of Science},
	title = {Synchrony of Sylvatic Dengue Isolations: A Multi-Host, Multi-Vector SIR Model of Dengue Virus Transmission in Senegal},
	year = {2012},
	month = {11},
	volume = {6},
	pages = {1-11}  
}

@article{moghadas2017asymptomatic,
	title={Asymptomatic transmission and the dynamics of Zika infection},
	author={Moghadas, Seyed M and Shoukat, Affan and Espindola, Aquino L and Pereira, Rafael S and Abdirizak, Fatima and Laskowski, Marek and Viboud, Cecile and Chowell, Gerardo},
	journal={Scientific Reports},
	volume={7},
	pages={5829},
	year={2017},
	publisher={Nature Publishing Group UK London}
}

@article{medeiros2021mathematical,
	title={A mathematical model for zoonotic transmission of malaria in the Atlantic Forest: Exploring the effects of variations in vector abundance and acrodendrophily},
	author={Medeiros-Sousa, Ant{\^o}nio Ralph and Laporta, Gabriel Zorello and Coutinho, Renato Mendes and Mucci, Luis Filipe and Marrelli, Mauro Toledo},
	journal={PLoS Neglected Tropical Diseases},
	volume={15},
	pages={e0008736},
	year={2021},
	publisher={Public Library of Science San Francisco, CA USA}
}

@article{sulaimon2025potential,
	title={The potential impacts of vector host species fidelity on zoonotic arbovirus transmission},
	author={Sulaimon, Tijani A and Wood, Anthony J and Bonsall, Michael B and Boots, Michael and Lord, Jennifer S},
	journal={PLoS Neglected Tropical Diseases},
	volume={19},
	pages={e0012196},
	year={2025},
	publisher={Public Library of Science San Francisco, CA USA}
}

@article{bouafou2024host,
	title={Host preference patterns in domestic and wild settings: Insights into Anopheles feeding behavior},
	author={Bouafou, Lemonde and Makanga, Boris K and Rahola, Nil and Bodd{\'e}, Marilou and Ngangu{\'e}, Marc F and Daron, Josquin and Berger, Audric and Mouillaud, Theo and Makunin, Alex and Korlevi{\'c}, Petra and others},
	journal={Evolutionary Applications},
	volume={17},
	pages={e13693},
	year={2024},
	publisher={Wiley Online Library}
}

@article{barbosa2018modelling,
	title={Modelling the impact of insecticide-based control interventions on the evolution of insecticide resistance and disease transmission},
	author={Barbosa, Susana and Kay, Katherine and Chitnis, Nakul and Hastings, Ian M},
	journal={Parasites \& Vectors},
	volume={11},
	pages={482},
	year={2018},
	publisher={Springer}
}

@article{kamiya2017epidemiological,
	title={Epidemiological consequences of immune sensitisation by pre-exposure to vector saliva},
	author={Kamiya, Tsukushi and Greischar, Megan A and Mideo, Nicole},
	journal={PLoS Neglected Tropical Diseases},
	volume={11},
	pages={e0005956},
	year={2017},
	publisher={Public Library of Science San Francisco, CA USA}
}

@article{ruan2008delayed,
	title={On the delayed Ross--Macdonald model for malaria transmission},
	author={Ruan, Shigui and Xiao, Dongmei and Beier, John C},
	journal={Bulletin of Mathematical Biology},
	volume={70},
	pages={1098--1114},
	year={2008},
	publisher={Springer}
}

@article{clancy2024extinction,
	title={Extinction in host--vector infection models and the role of heterogeneity},
	author={Clancy, Damian and Stewart, John JH},
	journal={Mathematical Biosciences},
	volume={367},
	pages={109108},
	year={2024},
	publisher={Elsevier}
}

@article{multerer2019modeling,
	title={Modeling the impact of sterile males on an Aedes aegypti population with optimal control},
	author={Multerer, Lea and Smith, Thomas and Chitnis, Nakul},
	journal={Mathematical Biosciences},
	volume={311},
	pages={91-102},
	year={2019},
	publisher={Elsevier}
}

@article{laperriere2011simulation,
	title={Simulation of the seasonal cycles of bird, equine and human West Nile virus cases},
	author={Laperriere, Vincent and Brugger, Katharina and Rubel, Franz},
	journal={Preventive Veterinary Medicine},
	volume={98},
	pages={99-110},
	year={2011},
	publisher={Elsevier}
}

@article{fairbanks2025quantifying,
  title={Quantifying vector diversion effects in zoonotic systems: A modelling framework for arbovirus transmission between reservoir and dead-end hosts},
  author={Fairbanks, Emma L and Baylis, Matthew and Daly, Janet M and Tildesley, Michael J},
  journal={PLoS Computational Biology},
  volume={21},
  pages={e1013359},
  year={2025},
  publisher={Public Library of Science San Francisco, CA USA}
   }

@article{tuschhoff2025heterogeneity,
  title={Heterogeneity in and correlation between host transmissibility and susceptibility can greatly impact epidemic dynamics},
  author={Tuschhoff, Beth M and Kennedy, David A},
  journal={Journal of Theoretical Biology},
  volume={611},
  pages={112186},
  year={2025},
  publisher={Elsevier}
}
 \bibliographystyle{elsarticle-num}

\end{document}